\documentclass{elsarticle}
\makeatletter
\def\ps@pprintTitle{%
 \let\@oddhead\@empty
 \let\@evenhead\@empty
 \def\@oddfoot{}%
 \let\@evenfoot\@oddfoot}

\usepackage[section]{placeins}
\usepackage{lineno,hyperref}
\biboptions{numbers,sort&compress}
\modulolinenumbers[5]

\journal{arXiv.org}

\usepackage{enumitem} 
\usepackage{multirow}
\usepackage{tabularx}
\newcolumntype{Y}{>{\centering\arraybackslash}X}

\newcommand{\para}[1]{\paragraph*{\textbf{#1}}}
\newcommand{\csentence}[1]{{\mathversion{bold}\bfseries #1}}%

\begin{document}

\begin{frontmatter}

\title{Academic Performance and Behavioral Patterns}

%% Group authors per affiliation:
%\author{Elsevier\fnref{myfootnote}}
%\address{Radarweg 29, Amsterdam}
%\fntext[myfootnote]{Since 1880.}

%% or include affiliations in footnotes:
\author[address1]{Valentin Kassarnig\corref{corauth}}
\ead{kassarnig@ist.tugraz.at}
\author[address2]{Enys Mones}
\author[address3,address4]{Andreas Bjerre-Nielsen}
\author[address2,address5]{Piotr Sapiezynski}
\author[address3,address4]{David Dreyer Lassen}
\author[address2,address4,address6]{Sune Lehmann}

\cortext[corauth]{Corresponding author. Institute of Software Technology, Graz University of Technology, Inffeldgasse 16B/II, 8010 Graz, Austria.}

\address[address1]{Institute of Software Technology, Graz University of Technology, Graz, Austria}
\address[address2]{Department of Applied Mathematics and Computer Science, Technical University of Denmark, Kgs. Lyngby, Denmark}
\address[address3]{Department of Economics, University of Copenhagen, Copenhagen, Denmark}
\address[address4]{Center for Social Data Science, University of Copenhagen, Copenhagen, Denmark}
\address[address5]{College of Information and Computer Science, Northeastern University,Boston,USA}
\address[address6]{The Niels Bohr Institute, University of Copenhagen, Copenhagen, Denmark}

\begin{abstract}
Identifying the factors that influence academic performance is an essential part of educational research.
Previous studies have documented the importance of personality traits, class attendance, and social network structure.
Because most of these analyses were based on a single behavioral aspect and/or small sample sizes, there is currently no quantification of the interplay of these factors.
Here, we study the academic performance among a cohort of 538 undergraduate students forming a single, densely connected social network. 
Our work is based on data collected using smartphones, which the students used as their primary phones for two years.
The availability of multi-channel data from a single population allows us to directly compare the explanatory power of individual and social characteristics.
We find that the most informative indicators of performance are based on social ties and that network indicators result in better model performance than individual characteristics (including both personality and class attendance).
We confirm earlier findings that class attendance is the most important predictor among individual characteristics.
Finally, our results suggest the presence of strong homophily and/or peer effects among university students. \end{abstract}

\end{frontmatter}

\section*{Introduction}
Since research on academic achievement began to emerge as a field in the 1960s, it has guided educational policies on admissions and dropout prevention~\cite{predictingacademic}.
Although much of the literature has focused on higher education, the knowledge obtained on behavioral phenomena observed in colleges and universities can potentially guide research on student behavior in primary and secondary schools.
A number of behavioral patterns have been linked to academic performance, such as time allocation~\cite{macan1990college}, active social ties~\cite{gavsevic2013choose}, sleep duration and sleep quality~\cite{curcio2006sleep}, or participation in sport activity~\cite{singh2012physical}.
Most of the existing studies, however, suffer from biases and limitations often associated with surveys and self-reports~\cite{van2008faking,junco2013comparing}, particularly when measuring social networks~\cite{kumbasar1994systematic, o2015social, freeman1992filling, bernard1984problem}.

Here we investigate the performance of 538 students within a novel dataset collected as part of the \textit{Copenhagen Network Study} (CNS), with data collection ongoing for more than two years~\cite{measuringlargescale}.
Due to the scale of the CNS, and the inclusion of directly observed data from smartphones in place of self-reports, we are able to mitigate some of the limitations encountered in existing `traditional' studies.
The strength of the CNS data is the high-resolution multi-channel measures for social interactions, including person-to-person proximity (using Bluetooth scans), calls and text messages, activity on online social networks (Facebook), and mobility traces. 

%Identifying students at risk of failing or dropping out at an early stage, is of particular interest from an administrative standpoint.
%In a practical scenario, these are the students that we might want to offer extra help.

The aim of our study was to better understand the impact of individual and network factors on our ability to distinguish between groups of students based on their performance. 
That is, we wanted to identify the ways in which low performers are significantly different from high performers and vice versa. 
We divide this goal into three specific objectives:
\begin{enumerate}[label=(\roman*),itemsep=0ex,parsep=1ex]
\item Identify individual and network factors that correlate with students' performances.
\item Analyze the importance of different sets of features for supervised learning models to classify students as low, moderate, or high performers. 
\item Investigate significant differences among performance groups for the most important individual and network features.
\end{enumerate}

\section*{Related Work}
\para{Individual behavior}
Through a variety of methods, a large number of studies have investigated the factors that determine academic performance.
Vandamme et al.~\cite{vandamme2007predicting} analyzed a broad range of individual characteristics concerning personal history, behavior, and perception.
Similarly, the \textit{StudentLife} study~\cite{wang2014studentlife} used smartphones to collect data on student activity, social behavior, personality, and mental health.
Both research groups observed correlations between performance and all feature categories, building a case that factors influencing academic performance are not limited to a single aspect of an individual's life.
Nghe et al.~\cite{nghe2007comparative} reframed the problem as a prediction task: using data to predict performance in a population of undergraduate and postgraduate students at two different institutions.
Using a wide range of features, they predicted GPA after third year with high accuracy. 
One of the features included GPA after the second year; in this work we show that even without the knowledge of past achievements it is possible to explain the students' performance levels to a large extent.
%in this work we show that it is possible to make accurate predictions without the knowledge of past performance.
Furthermore, prior research has emphasized the positive influence of attending classes~\cite{buckalew1986relationship, marburger2006does, chen2008class, crede2010class}.
The study by Crede et al.~\cite{crede2010class} concludes that attendance is the most accurate known predictor of academic performance; see \cite{attendance2017companionpaper} for a more detailed analysis of the impact of class attendance on academic performance based on the CNS data.

Cao et al.~\cite{cao2017orderness} analyzed behavioral data from the digital records of nearly 19,000 students' smart cards, such as entering and leaving the library, having a meal in the cafeteria, or taking a shower in the dormitory. 
They conclude that the students' orderness (regularity of daily activities) is a strong predictor of academic performance.
Our approach shares some similarities with~\cite{cao2017orderness}, but the key difference is that we have investigated not only individual behavior but also the students' social environment.

\para{Individual traits}
A large body of research at the intersection of psychology and education investigated the relationship between personality and performance, as pioneered by~\cite{prociuk1974locus}. Many personality traits were found to be linked to academic success: Among the dimensions of the well-studied Big-Five Inventory~\cite{goldberg1993structure} \textit{Conscientiousness} (positive) and \textit{Neuroticism} (negative) displayed the strongest correlation with academic performance~\cite{dollinger1991personality, goff1992personality, rothstein1994personality, wolfe1995personality, de1996personality, paunonen1998hierarchical, busato2000intellectual, paunonen2001big, gray2002general, lievens2002medical, bauer2003effect, chamorro2003personalitya, chamorro2003personalityb, diseth2003personality, farsides2003individual, furnham2002personality, lounsbury2003intelligence, phillips2003personality, duff2004relationship, furnham2004personality, hair2006role, conard2006aptitude, barchard2003does, langford2003one, oswald2004developing, leong2005sense, ridgell2004predicting, komarraju2009role, noftle2007personality}. 
The other three dimensions showed only very weak or no correlation.
%; \textit{Extraversion} was rather positively and \textit{Openness} rather negatively correlated. 
Furthermore, the characteristics \textit{Self Esteem}~\cite{lane2004self}, \textit{Satisfaction with Life}~\cite{lepp2014relationship, chow2005life}, and \textit{Positive Affect Schedule}~\cite{saklofske2012relationships} were also found to be positively correlated, while \textit{Stress}~\cite{stewart1999prospective, akgun2003learned}, \textit{Depression}~\cite{haines1996effects, leach2009relationship, owens2012anxiety}, and  \textit{Locus of Control}~\cite{lepp2014relationship, chow2005life} showed a negative effect on academic achievements.

\para{Online Social Media}
Only a few prior studies have investigated the impact of social media activity on academic performance, despite the growing availability of such data and undisputed presence of these media in our daily lives.
The majority of existing studies found a decrease in academic performance with increasing time spent on social media~\cite{maqableh2015impact, al2015social, al2014relationship, karpinski2013exploration, paul2012effect, junco2012relationship, jacobsen2011wired, kirschner2010facebook}. 
However, not all studies confirm this result. 
In some studies, time spent on social media was found to be unrelated to academic performance~\cite{pasek2009facebook, ainin2015facebook} or even a had positive effect on performance~\cite{kolek2008online, tayseer2014social}.

\para{Social Interactions}
There is a growing interest in the relationship between social interactions (especially online social interactions) and academic performance~\cite{sacerdote2001peer, zimmerman2003peer, stinebrickner2006can, carrell2013natural, vitale2016examining, smirnov2016formation, poldin2013social, mayer2008old, yuan2006focused, rizzuto2009s, gavsevic2013choose, tomas2015influence, sparrowe2001social, smith2007psst,hommes2012visualising, baldwin1997social, yang2003effects, cho2007social, thomas2000ties, johnson1984structuring}. 
In the relevant literature there exist two dominant approaches.
The first approach focuses on the relation between own performance and that of peers~\cite{sacerdote2001peer, zimmerman2003peer, stinebrickner2006can, carrell2013natural, poldin2013social, vitale2016examining, smirnov2016formation, mayer2008old}, based on a hypothesis of similarity in peer achievement.
The similarity between pairs of individuals connected via social ties are attributed to various aspects: selection into friendships by similarity (i.e., homophily); influence by social peers (also know as peer effect); and correlated shocks (e.g., being exposed to the same teacher). 
As noted by \cite{manski1993identification,sacerdote2001peer} the issue of separating these effects is inherently difficult.
The second approach emphasizes the positive influence of having a central position in the social network between students~\cite{baldwin1997social, sparrowe2001social, yang2003effects, cho2007social, smith2007psst, hommes2012visualising}.
The majority of results in the existing research which measure social networks are, however, based on self-reports and therefore subject to various biases~\cite{kumbasar1994systematic, o2015social, freeman1992filling, bernard1984problem} that are in many ways mitigated by using smartphones to measure the social network~\cite{eagle2008mobile}. 
However, it should be noted that surveys and observational studies often measure very different aspects of reality. 
For instance, in the case of assessing tie strengths, observational studies may be more accurate in quantifying duration and frequency variables of a relationship, while surveys can provide qualitative insights into depth and intimacy~\cite{marsden1984measuring, newcomb1995children}.

%%%%%%%%%% METHODS %%%%%%%%%%
\section*{Materials and Methods}
\subsection*{Data collection \& preprocessing}
Results presented in this paper are based on the data collected in the \textit{Copenhagen Network Study} (CNS)~\cite{measuringlargescale}.
In the CNS, dedicated smartphones where handed out to students at the Technical University of Denmark (DTU) and used as their primary phones for two years. 
During this period various data types were recorded: Bluetooth scans, call and text message meta data, Facebook activity logs, and mobility traces. 
Additionally, participating students answered a survey on personality at the beginning of the study.
Due to the possibility to exit the experiment at any given point, the number of participants varied over time. 
We investigate the data from 538 undergraduate students for whom we have complete data.

%Bluetooth + Facebook data
The raw data records are cleaned and transformed to meaningful information before the analysis.
Bluetooth scans are used to estimate person-to-person interactions corresponding to a physical distance of up to 10\,m (30\,ft) between participants. While physical proximity is not a perfect proxy for person-to-person interactions, there is evidence that the proximity interactions are predictive of friendship in online social networks and communication using phone calls and text messages~\cite{sekara2014strength,sapiezynski2017inferring,eagle2009inferring}.

%After constructing proximity links, data is binned in 15 minutes time bins.
Facebook data was obtained via the Facebook Graph API, and contains both static friendship connections as well as various interactions on the social network. All types of interactions are treated equally. 
Private messages, however, are unavailable since they cannot be obtained from Facebook using the official Graph API.

%Location data + attendance data
The location data on the smartphones has varying accuracy depending on the providing sensor. The accuracy of the collected position can vary between a few meters for GPS locations, to hundreds of meters for cell tower location. 
We group the location data into 15-minute bins and use the median location of all data points with an accuracy below 80\,m.
In order to compute attendance we combined the smartphone locations with the person-to-person proximity obtained from Bluetooth scans. 
A detailed description of the method can be found in a companion paper~\cite{attendance2017companionpaper}.

%Networks
We considered social interactions of five different channels: proximity, Facebook (friendships + interactions), calls, and text messages. For each channel we created a network to model the social relations.  
Note that these models are based only on the interactions among participants of the CNS. 
Interactions with any people outside the study were not considered. 
Importantly, for the proximity networks we excluded all meetings that took place during class time in order to eliminate effects caused by class co-attendance. Section B in the Supporting Information discusses further details of the creation of these network models.
In the remainder of this paper, the direct neighbors in those networks are refereed to as `peers'. 

%We use the terms peers, friends, and contacts interchangeably throughout this paper to describe the direct neighbors in the social network graph. 

\begin{figure}[!h]
\centering
	\includegraphics[width=\textwidth]{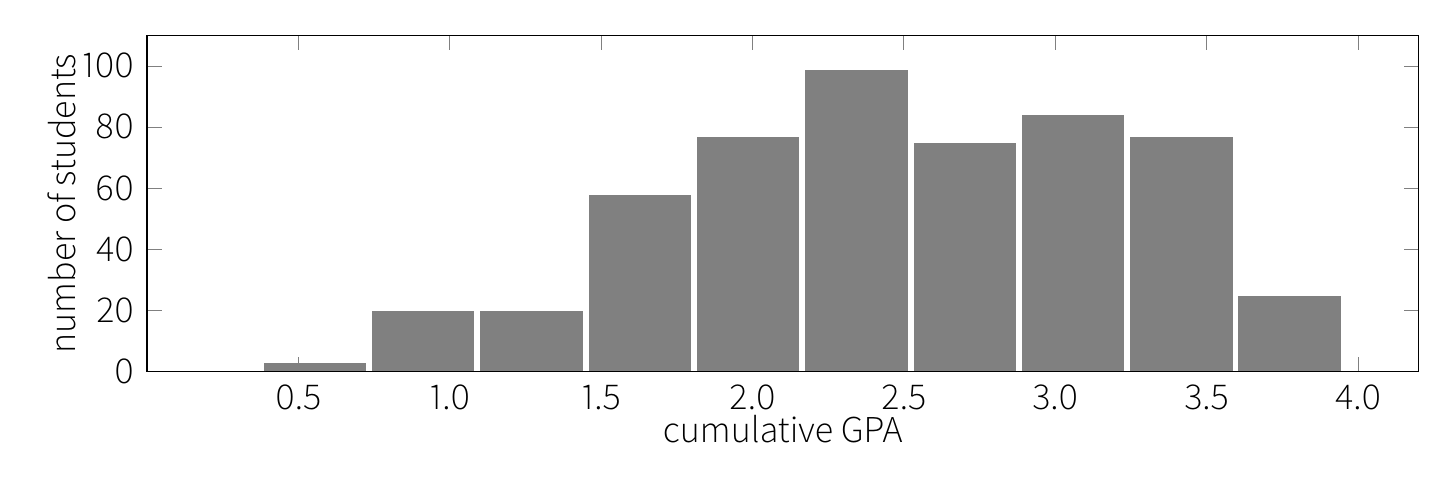}
	\caption{\csentence{Distribution of cumulative GPAs.} Distribution of 538 cumulative GPAs. The histogram shows a left-skewed distribution with a mean GPA of 2.5.}
	\label{fig:gpa_histogram}
\end{figure}

%Grades + Population
The students' course grades were provided by DTU administration.
Only courses using the Danish 7-point grading scale were considered. 
This scale consists of the grades 12, 10, 7, 4, 02, 00, and -3 with 12 being the best grade and 00 and -3 indicating that the student failed. 
The positive weighted mean grades (term or cumulative) were converted to the standard GPA scale ranging from 4.0 (best) to 0.0 (worst). 
Every negative mean grade was set to 0.0. 
Only students attending at least three courses were considered. Fig.~\ref{fig:gpa_histogram} illustrates the distribution of the 538 cumulative GPAs. 
It shows a left-skewed distribution with a mean GPA of 2.5. More information about the student population can be found in Section A of the Supporting Information.

%Bootstrapping
In order to increase the stability of the results we applied bootstrap resampling. Analyses were performed on 100 bootstrap samples, where each has the same size as the original sample. 
We report as results the mean of the bootstrap analyses with approximated standard errors described by the \emph{Standard Error of the Mean}.

\subsection*{Feature sets}
To account for the different explanatory power of the individual and network aspects, we constructed four feature sets, each representing a certain aspect of life and corresponding to a specific level of information: \emph{personality}, \emph{individual}, \emph{network} and \emph{combined}. 

\para{Personality features}
The personality features contain 16 individual personality traits obtained from questionnaires that the study participants had to fill in before receiving a phone.

\para{Individual features}
The individual feature set combines the 16 personality traits with behavioral and personal variables. 
Behavioral variables include average class attendance and the Facebook activity level (log of average number of posts per week). 
In terms of personal information, we added the students' gender and their study year to the feature set. Information about the sociological background of the students was not available to us.

\para{Network features}
For the network features we consider metrics from five different networks, each based on a different channel (texts, calls, proximity, Facebook interactions, and Facebook friendships).
Despite the large number of possible features to extract from networks, we considered only the metrics that follow the main approaches found in the literature, such as the mean GPA of peers, centrality, and the fraction of low and high performing peers.
However, further aspects, such as deviation, skewness, or entropy of peers' GPAs, would undoubtedly be interesting for future investigations.

The structure of the interaction networks provide further insight into how students' position in their social environment is correlated with performance. 
Therefore, we evaluated different centrality measures\footnote{Details on the evaluation can be found in Section~C of the Supporting Information.}. 
Overall, the degree centrality displayed the strongest correlation and was therefore used as feature in our analyses. 
%number of contacts (degree) and eigenvector centrality.
%Degree centrality quantifies the number of direct contacts a person in the network has.
%Eigenvector centrality is related to flows in the social network and individuals of high centrality are regarded as influencers in case of a spreading dynamics (outbreak, rumors, etc.)~\cite{fowler2008dynamic}.

\para{Combined features}
The combined feature set contains all 20 individual features and all 20 network features yielding a total of 40 features.
See Table~\ref{tab:feature_sets} for a complete list of features in each category. More details including descriptive statistics can be found in Section E of the Supporting Information.

\subsection*{Approach}
We use machine learning techniques to evaluate the importance of different factors on the academic performance of students. 
Specifically, we create supervised learning models and evaluate their performance on classifying students as low, moderate, or high performers. This framework allows us to compare our results to related work, in particular, the works by Vandamme et al.~\cite{vandamme2007predicting} and Nghe et al.~\cite{nghe2007comparative}. 
Furthermore, this approach makes it easier to detect significant differences between the individual performance groups.
In contrast to classical statistical modeling with test of significance, machine learning uses a hypothesis-free approach that allows us to model complex interactions driven by the data~\cite{valletta2017applications}.
We evaluate the model performance based on the mean classification accuracy of 100 independent 10-fold cross-validations.

%Model selection was carried out by a 10-fold cross-validation, and due to variations in the different realizations, we present the average performance metrics of 100 independent cross-validations.

A key point to emphasize here is that while classifying students' performance levels based on current behavior might be useful in a practical context (for example to identify students in need of extra support), it is not our primary reason for using machine learning in the current study.
Rather, we use machine learning as a tool for ranking and comparing features.
That is, the more predictive a given feature is, the more important it is for describing performance.
By training our models on features arising from many categories, previously only studied independently, we can begin to understand their relative importance, as well as their interplay in terms of academic performance.

%%%%%%%%%% RESULTS %%%%%%%%%%
\section*{Results}

The following results are reported in three stages. First, we perform an ANOVA F-test on all features to identify the most important features for dividing students into performance groups. 
Then we utilize supervised learning models to investigate the importance and interplay of the different feature categories.
Based on the results of the first two stages, we then conduct an in-depth analysis of the most expressive impact factors of each category. 
Our primary focus is on the social behavioral features which have only been considered to a limited extent in previous studies.

\subsection*{Analysis of variance}

\begin{figure}[!h]
\centering
	\includegraphics[width=\textwidth]{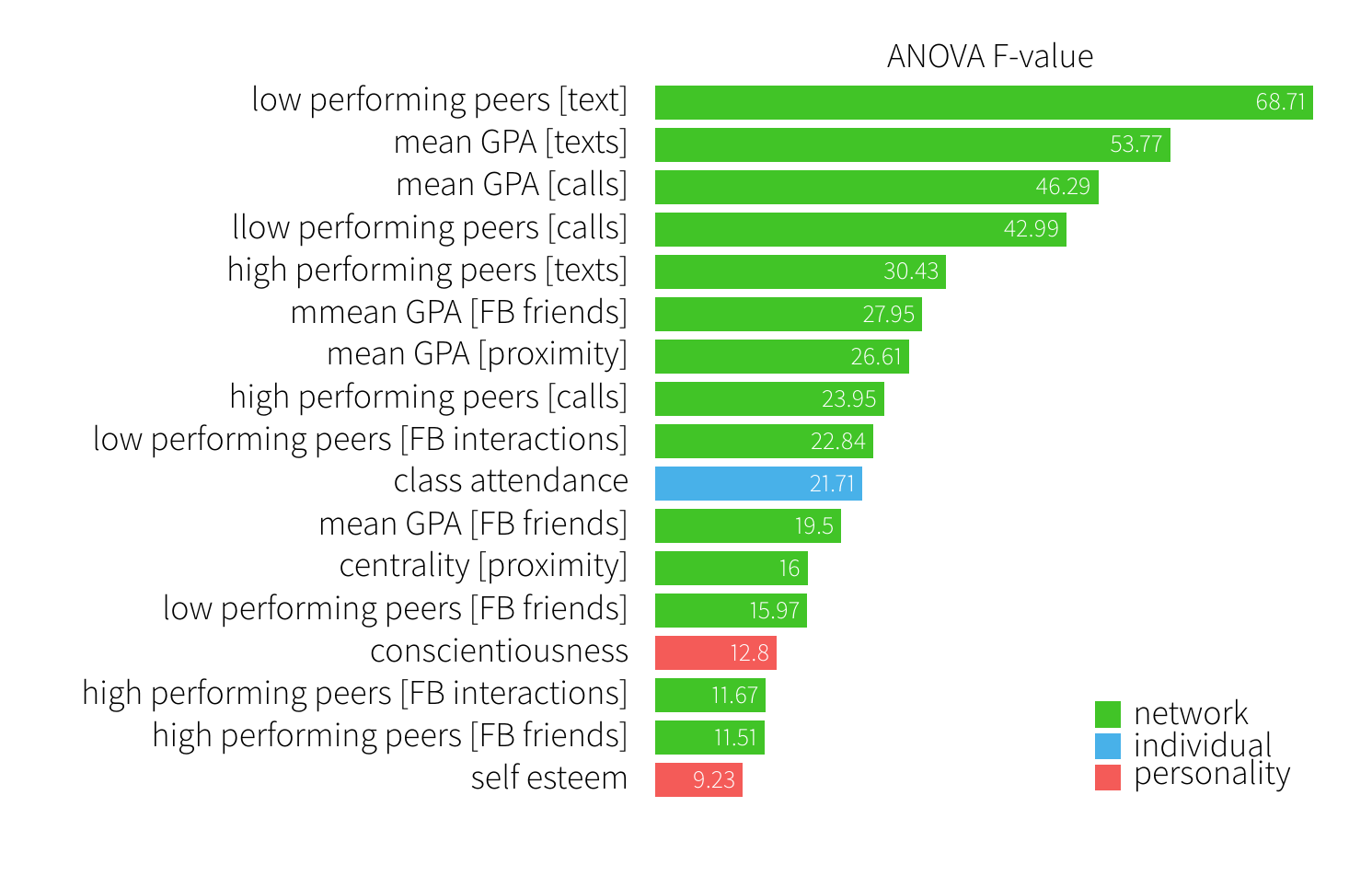}
	\caption{\csentence{Feature importance ranking.} Results from ANOVA F-test for 3-class classification. Features which did not achieve sufficient significance ($p \geq .001$) are omitted.}
	\label{fig:feature_importance}
\end{figure}

Fig.~\ref{fig:feature_importance} shows the feature importance for features achieving significance of $p < .001$ obtained from an ANOVA F-test.\footnote{Note that F-test should not be interpreted literally here, as the assumption of identical independent draws of errors is likely to violated due to correlation of errors in the network. Rather, we use it only as a guide to select features.}
Although all feature categories are correlated with academic performance, the result indicates that features which describe the social networks of students have the highest explanatory power.
In general, network properties dominate the results with more than half of the significant features corresponding to this category. 
A potential explanation for the high impact of social relations is that the network connections may act as a proxy for previous performance, since the network features include information on the grades of others.
The fraction of low performing peers as well as the mean GPA of peers contacted over text messages and calls display the highest explanatory power.\footnote{The reliability of this observation has been validated by a permutation test -- see Section D of the Supporting Information}
Class attendance proves to be the most important individual feature and moreover, overall the most important one if we had no information on anyone's grades. 
Centrality in the proximity network is also found to be a significant descriptor with moderate importance.
Among personality traits, only self-esteem and conscientiousness have significant explanatory power.

\subsection*{Supervised learning}
In order to better understand the importance and interplay of different factors on the academic performance we utilized supervised learning techniques. We created models based on the different feature sets to classify the students as low, moderate, and high performers according to their GPAs. Each of those three groups contains the same number of students, corresponding to a baseline accuracy of 33.33\%.

\begin{figure}[!h]
	\centering
	\includegraphics[width=\textwidth]{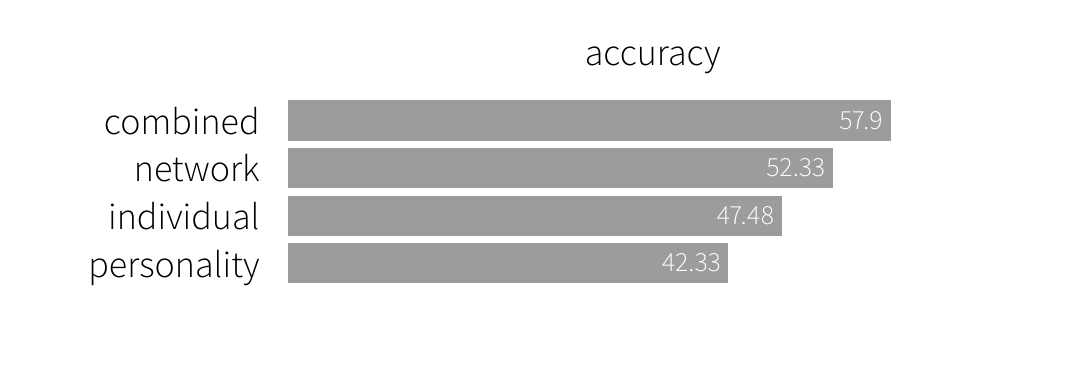}
	\caption{\csentence{Model performances on the different feature sets.}
		Bars show the classification accuracy of the different LDA models.}
	\label{fig:model-performance}
\end{figure}

\begin{figure}[!h]
\centering
	\includegraphics[width=0.75\textwidth]{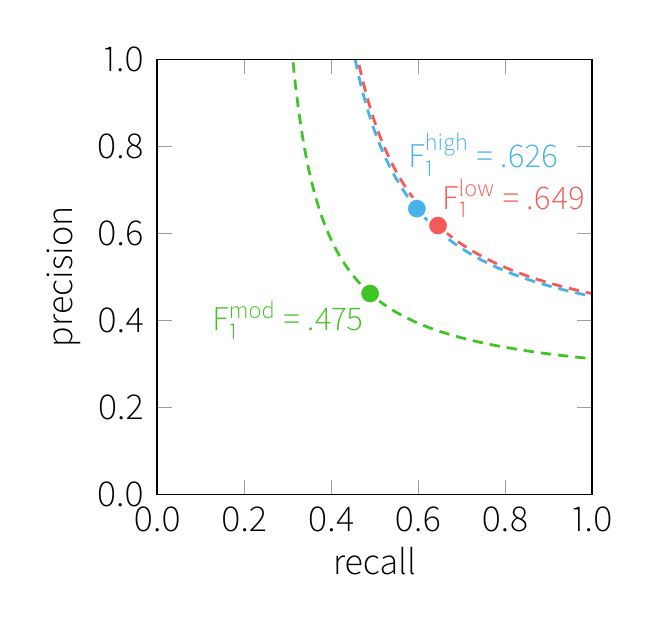}
	\caption{\csentence{Precision-recall curve.}
		Dots represent the model performance in the low (red), moderate (green) and high (blue) performer classes.
		Dashed lines mark the profile of constant $F_1$ corresponding to the measured values for the specific class.}
	\label{fig:precision-recall}
\end{figure}

We use Linear Discriminant Analysis (LDA) to find an optimal model that separates the three performance classes. 
Fig.~\ref{fig:model-performance} illustrates the mean results of 100 independent 10-fold cross-validations. The results show that the LDA model solely based on personality features exceeds the baseline performance by about 9 pps. Adding the four additional individual features (behavior + background info) improves the model's performance by further 5.2 pps. Using network features instead of individual features results in a performance of about 19 pps above baseline. Combining individual and network features yields a superior model with about 57.9\% accuracy; roughly 25 pps above baseline.
Fig.~\ref{fig:precision-recall} shows its achieved in-class precision and recall values along with the corresponding $F_1$ values.
As the results indicate, once the GPA class is provided, the model has high predictive power among the low and high performers (compared to that of the moderate performers) with $F_1$ values of .632 and .626, respectively.

% To supporting information??
%Identifying students at risk of failing or dropping out at an early stage, is of particular interest from an administrative standpoint.
%In a practical scenario, these are the students that we might want to offer extra help.
%Modeling the question of identifying these students as a binary classification allows us to identify 77.4\% of the at-risk students (bottom third) with a precision of 58.3\%. 
%This corresponds to a $F_1$ score of .666. 
%On the other hand, non-risk students were correctly classified as such with a precision of 86.5\% ($F_1$ = .788). 
%These results underline the predictive power of the observed variables, in particular in the application of ascertaining at-risk students.

\subsection*{Feature analysis}

\para{Individual behavior}
Among the considered individual effects, class attendance was found to have the highest impact on academic performance.
A correlation coefficient of $r_S = .294$ for cumulative GPAs was determined ($p < .001$).
An in-depth analysis of the observed class attendance patterns along with a detailed description of the method to measure attendance in the CNS dataset is discussed in~\cite{attendance2017companionpaper}.

\begin{figure}[!h]
	\centering
	\includegraphics[width=\textwidth]{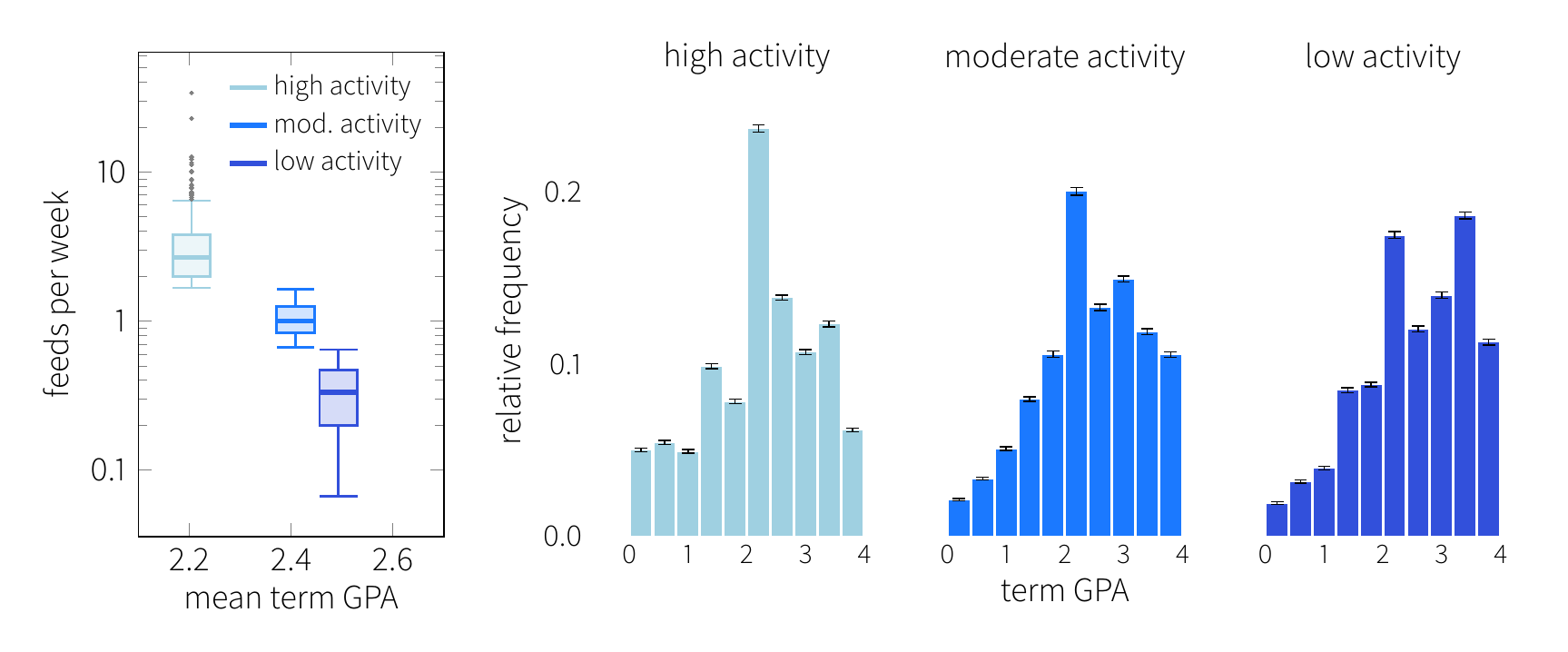}
	\caption{\csentence{Facebook usage and performance in the tertiles.}
		(a) Division of students into three groups of equal size according to their active Facebook updates.
		Each box represents a single tertile, width corresponds to the span of Facebook activity in the specific group and the x-position shows the mean term GPA.
		(b) Grade distribution inside each Facebook activity class.}
	\label{fig:FB_activity_dist}
\end{figure}

The Facebook activity level measures the average number of published posts. Since the activity levels change significantly over time we consider each semester separately and use the corresponding term GPAs as measure for academic performance. This gives us up to four data points per student (one for each semester of the data collection period) for this analysis. In Fig.~\ref{fig:FB_activity_dist} students are divided into three groups of equal size according to their activity levels. As Fig.~\ref{fig:FB_activity_dist}a shows, the distribution of posts among students is heavy-tailed and is described by the vast majority of the students having less than 3 posts in a typical week.
The distribution of term GPA values in the different tertiles reveals that, on average, students with lower activity perform better (see Fig.~\ref{fig:FB_activity_dist}b).
To statistically evaluate the variation in the distribution over the different tertiles, we performed a Kruskal--Wallis H-test. 
This test rejected the global null hypothesis with $p<.001$ that the medians of the groups are all equal. 
A follow-up Dunn multiple comparison test with Bonferroni correction revealed pair-wise differences among the tertiles: all pairs are significantly different from each other ($p<.001$). 
Thus, groups with different levels of Facebook activity have significantly different academic performances.

\begin{figure}[!h]
	\centering
	\includegraphics[width=\textwidth]{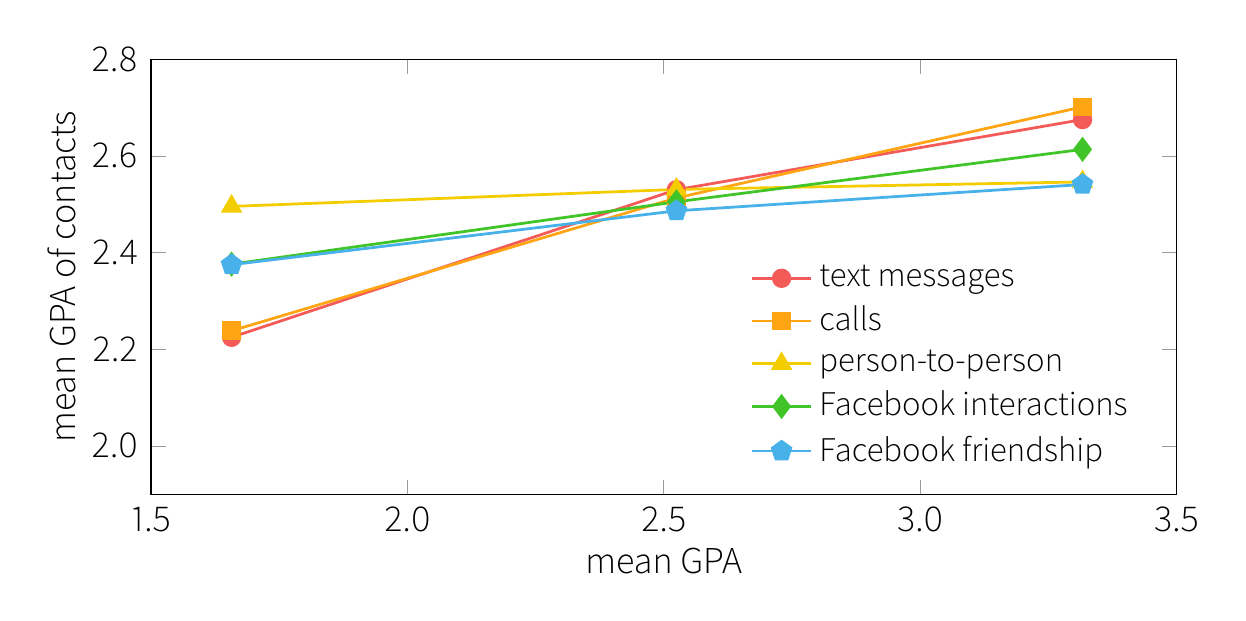}
	\caption{\csentence{Similarity in academic performance for social ties.}
		Curves show the mean GPAs of every performance group and their peers from different communication channels.}
	\label{fig:homophily}
\end{figure}

\para{Social interactions}
Based on the results presented in Fig.~\ref{fig:feature_importance} and Fig.~\ref{fig:model-performance} we conclude that a student's performance can be accurately inferred from the achievements of their peers. This effect was consistently observed across different communication and interaction channels, as shown in Fig.~\ref{fig:homophily}. There, each channel is represented by a separate line illustrating the mean correlation of the members of each performance group and their respective peers.
We can observe that regardless of the channel considered, each curve shows a strong increasing trend. This is further quantified in Table~\ref{tab:corr_GPA_friends} which displays the corresponding correlation coefficients on the individual level.
The most pronounced effect is observed for calls and text messages, which are considered to be proxies for strong social ties because this type of connection requires effort to initiate and maintain~\cite{van2010ll}.

Interestingly, these channels are not dominant in the case of centrality measures. 
Here, proximity interactions displayed the strongest correlation among all channels. 
However, we found weak to moderate positive correlations in all social networks, in agreement with the existing literature~\cite{baldwin1997social,yang2003effects,cho2007social,smith2007psst,hommes2012visualising,sparrowe2001social}.

\begin{figure}[!h]
	\centering
	\includegraphics[width=0.75\textwidth]{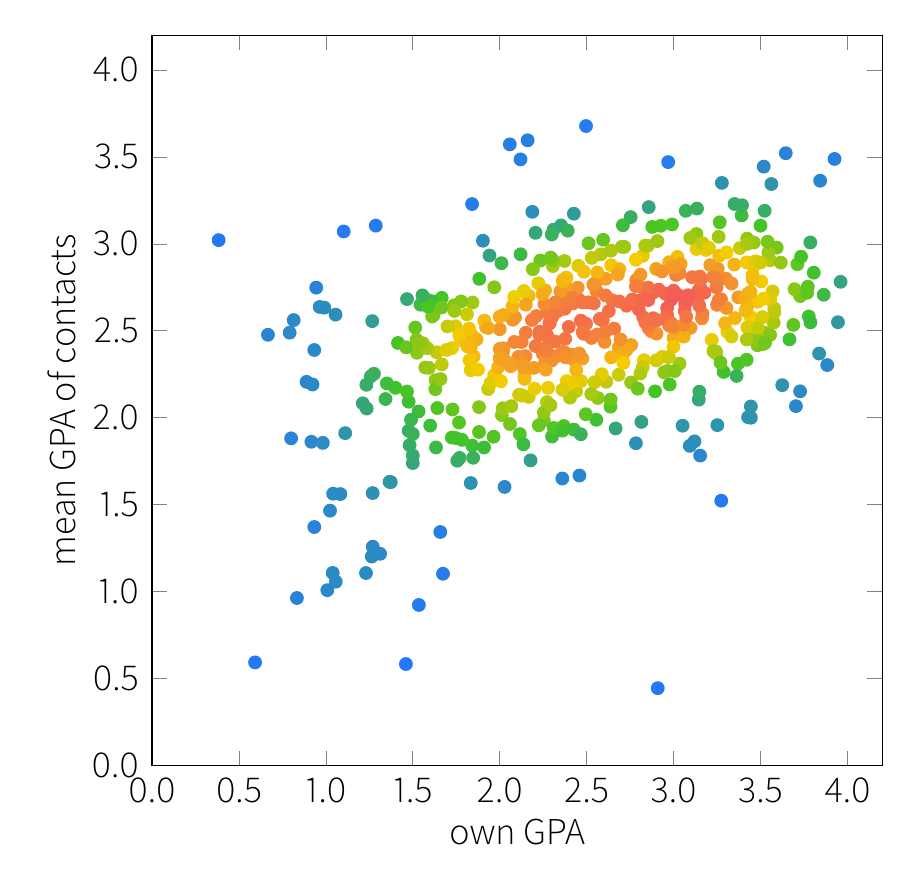}
	\caption{\csentence{Correlation between performance of strong peers.}
		For each student, we show their cumulative GPA versus the mean GPA of their peers obtained by their text messages.
		Color denotes density of points in arbitrary units.}
	\label{fig:text_scatter}
\end{figure}

\begin{figure}[!h]
	\centering
	\includegraphics[width=\textwidth]{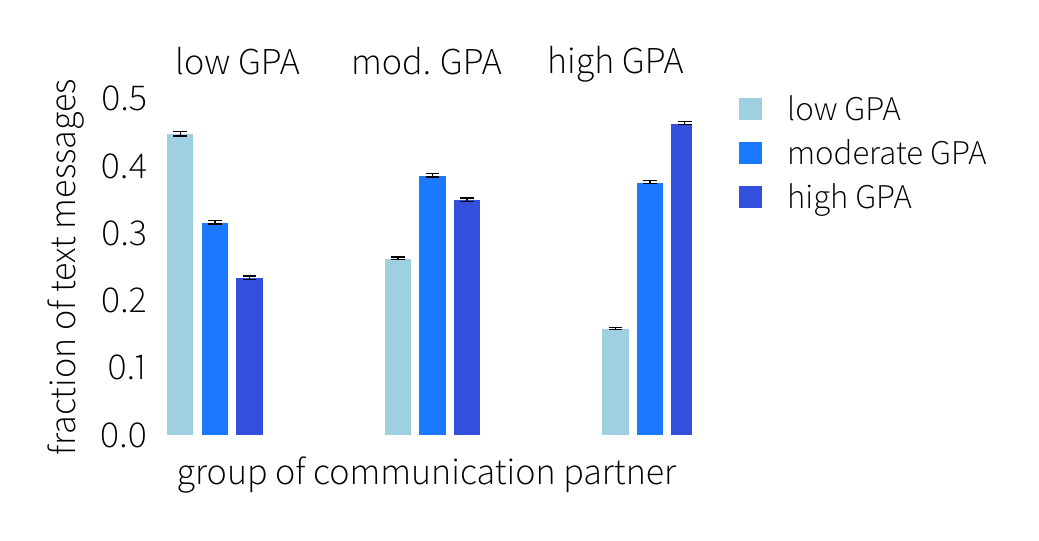}
	\caption{\csentence{Own academic performance and peers' academic performance.}
		Each histogram displays how students distribute their text messages exchanged with others over the various performance groups.
		Groups are defined by tertiles based on their cumulative GPA.}
	\label{fig:homophily_texts}
\end{figure}

We further assessed the validity of pairwise similarity in the network by focusing exclusively on social ties based on text messages.
Fig.~\ref{fig:text_scatter} shows a scatter plot of the correlation between the own GPA and mean GPA of the texting peers for every student in the dataset.
Once again, we observe a clear linear trend; the trend is especially strong in the region where the majority of the students is located (GPAs in the range between 2 and 3).
In Fig.~\ref{fig:homophily_texts} we divided the population into tertiles based on the GPA and calculated the fraction of text messages exchanged with members of the different groups. 
Beyond the correlation, we can see that the students' communication in each group is dominated by members of the same group.
This observation further underlines the importance of the social environment for academic success.

%It should be noted that the proximity networks include all connections, even those formed on campus, and during lectures, which may account for some of the observed correlations.

%%%%%%%%%% DISCUSSION %%%%%%%%%%
\section*{Discussion}
For the participants of the CNS, we found that the peers' academic performance has a strong explanatory power for academic performance of individuals. 
We observed this effect across different channels of social interactions with calls and text messages showing the strongest correlations, further emphasizing the phenomena.
As mentioned in the literature review, this effect could be caused by either peer effects (adaption) or homophily (selection).
It should be noted that GPA information is used here as target and, in aggregated form, also as network feature. This allows us to analyze and understand the relationships among peers; but should be taken into account when framing the problem as prediction task.

We found network centrality to have a positive correlation with academic performance, in agreement with the literature~\cite{baldwin1997social, yang2003effects, cho2007social, smith2007psst, hommes2012visualising, sparrowe2001social}.
However, among all types of interaction networks, only proximity networks exhibited a strong effect.
A possible limitation in measuring centrality is that the mere physical proximity of two individuals does not necessarily involve direct communication. 
Nevertheless, it is reasonable to expect an increased level of information exchange in a group of individuals if they are in close proximity, which was the case in our dataset.\footnote{The CNS uses (thresholded) Bluetooth visibility as an indicator of person-to-person proximity }

Consistent with findings in existing literature, we found that class attendance showed the strongest correlation with academic performance when we consider only individual effects~\cite{stanca2006effects, buckalew1986relationship, chen2008class, crede2010class, van1992class, brocato1989much, gump2005cost, lin2006cumulative}.
We also found that Facebook activity has a negative relation to academic performance -- also in agreement with the majority of the studies that investigated Facebook and social media usage~\cite{maqableh2015impact, al2015social, al2014relationship, karpinski2013exploration, paul2012effect, junco2012relationship, jacobsen2011wired, kirschner2010facebook}.
We note, however, that our the data is limited to Facebook activities such as posting a status update or uploading a picture etc, and that we have no information regarding `passive' Facebook usage, such as scrolling and reading.
Also, our data does not include direct messages which may constitute a relevant fraction of communications performed via the social network site.

The analysis of the different personality traits revealed that two characteristics, namely conscientiousness and self-esteem, have considerable explanatory power for academic success.
These two traits reached a correlation coefficient between 0.2 and 0.3 corresponding to the upper limit achievable for any correlation with a personality trait, according to Mischel~\cite{mischel2013personality}. The impact of other investigated characteristics could not be confirmed with proper significance.
These results agree with existing literature~\cite{dollinger1991personality, goff1992personality, rothstein1994personality, wolfe1995personality, de1996personality, paunonen1998hierarchical, busato2000intellectual, paunonen2001big, gray2002general, lievens2002medical, bauer2003effect, chamorro2003personalitya, chamorro2003personalityb, diseth2003personality, farsides2003individual, furnham2002personality, lounsbury2003intelligence, phillips2003personality, duff2004relationship, furnham2004personality, hair2006role, conard2006aptitude, barchard2003does, langford2003one, oswald2004developing, leong2005sense, ridgell2004predicting,komarraju2009role, noftle2007personality, lane2004self}.

In the supervised learning experiment we achieved a classification accuracy of around 25 percentage points above baseline, a result similar to that of Vandamme et al.~\cite{vandamme2007predicting}
While the classification accuracy is similar, comparing our results with theirs is difficult because of the very different feature sets and experimental setups. 
Vandamme et al.~\cite{vandamme2007predicting} use nearly ten times as many features to build a model as we did.
In addition, the accuracy of Vandamme et al.~\cite{vandamme2007predicting} is driven by using prior achievement (grades), which is known to be a strong predictor of performance (e.g.~due to persistence of skill and motivation).
We note here that a potential reason for the similarity in performance to Vandamme et al.~\cite{vandamme2007predicting} could be that the network features used in our study include the grades of others in the network. 
Thus, if the network homophily with respect to academic performance is sufficiently strong, the average performance of others could serve as a proxy for each individual's academic achievements. 

Networks originating from different channels were treated separately because each network provides different information. 
For future studies it could be interesting to combine them and create multiplex network models which capture interactions across multiple channels and provide more information about the actual tie strength. 

In summary, our findings---together with the results in the literature---emphasize that there is a considerable dependence of academic performance on personality and social environment. 
This experiment is by no means an attempt to be exhaustive of the possibilities for impact factors. 
Rather, we hope that this demonstration will stir interest to further study the impact of the social environment on academic success, as well as the interplay of individual and network factors. 
%However, there are also other factors that play a role which we were unable to capture.

\subsection*{Limitations}

Although we utilized wider and more detailed data than most other studies, our approach also has important limitations which need to be taken into account.
First, we only observed students from a single, technical, Danish university. 
For this reason, the findings may not be generalizable to students at other institutions, of other academic disciplines or with other demographics. 
Furthermore, only a subset of all the students at DTU participated in our study -- for first year students the rate was around 40\%.
Although we observed a high degree of variation with respect to behavioral and network measures as well as academic performance, our sample may not be representative of the whole student population. Our measures of ego-networks and model estimates reflect only the smaller (and not closed) community of students in the CNS within the larger population of students.

%Yet, within our dataset we observed a high degree of variation with respect to behavioral and network measures as well as academic performance. 
%Thus, we argue that our results are valid indicators for the trends present in the larger student population.

Although direct measures overcome a lot of the limitations of surveys and self-reports, they continue to be affected by standard concerns over observational data, including selection bias, information bias, and confounding~\cite{hill2000bias}. 
In particular, confounding plays a big role in our study as there are many factors that we were unable to capture but provenly affect the academic performance directly or interplay with other observed factors. For instance, many socio-economic variables have been identified as good predictors for academic achievements~\cite{deberard2004predictors,cohn2004determinants,white1982relation,sirin2005socioeconomic} but unfortunately such data was not available to us.
There was also some tendency of selection into the study as the average student in the study tends to achieve higher grades than non-participants~\cite{bjerre2017dynamics}. 
Furthermore, investigations on the CNS data have revealed, that findings differ slightly for men and women~\cite{sapiezynski2017academic}.

% removing students at random due to technical reasons (missing data)
 
Social network observations were limited to phone calls/texts, meetings, and Facebook activities. 
Although these are arguably some of the most important means of communication, some students may communicate via other smartphone apps. 
Our method of inferring attendance is also subject to some noise (as thoroughly discussed in~\cite{attendance2017companionpaper}). Furthermore, it does not imply in-class participation nor attention to the taught material.

Although we have identified many factors that correlate with academic performance, we make no claims regarding causality. 
The question of establishing causality from purely observational data is far from trivial.
Thus, while being beyond the scope of this work we consider this question as promising and interesting for future research.

\section*{Competing interests}
The authors declare that they have no competing interests.

\section*{Funding}
This work was supported by the Villum Foundation, the Danish Council for Independent Research, University of Copenhagen (via the UCPH-2016 grant Social Fabric and The Center for Social Data Science) and Economic Policy Research Network (EPRN).

\section*{Author's contributions}
All authors contributed equally to this work.

\section*{Acknowledgments}
Due to privacy implications we cannot share data but researchers are welcome to visit and work under our supervision.

%% BioMed_Central_Bib_Style_v1.01

\newcommand{\BMCxmlcomment}[1]{}

\BMCxmlcomment{

<refgrp>

<bibl id="B1">
  <title><p>The prediction of academic performance</p></title>
  <aug>
    <au><snm>Lavin</snm><fnm>DE</fnm></au>
  </aug>
  <publisher>New York City, USA: Russell Sage Foundation</publisher>
  <pubdate>1965</pubdate>
</bibl>

<bibl id="B2">
  <title><p>College students' time management: Correlations with academic
  performance and stress.</p></title>
  <aug>
    <au><snm>Macan</snm><fnm>TH</fnm></au>
    <au><snm>Shahani</snm><fnm>C</fnm></au>
    <au><snm>Dipboye</snm><fnm>RL</fnm></au>
    <au><snm>Phillips</snm><fnm>AP</fnm></au>
  </aug>
  <source>Journal of educational psychology</source>
  <publisher>American Psychological Association</publisher>
  <pubdate>1990</pubdate>
  <volume>82</volume>
  <issue>4</issue>
  <fpage>760</fpage>
</bibl>

<bibl id="B3">
  <title><p>“Choose your classmates, your GPA is at stake!” The association
  of cross-class social ties and academic performance</p></title>
  <aug>
    <au><snm>Ga{\v{s}}evi{\'c}</snm><fnm>D</fnm></au>
    <au><snm>Zouaq</snm><fnm>A</fnm></au>
    <au><snm>Janzen</snm><fnm>R</fnm></au>
  </aug>
  <source>American Behavioral Scientist</source>
  <publisher>Sage Publications Sage CA: Los Angeles, CA</publisher>
  <pubdate>2013</pubdate>
  <volume>57</volume>
  <issue>10</issue>
  <fpage>1460</fpage>
  <lpage>-1479</lpage>
</bibl>

<bibl id="B4">
  <title><p>Sleep loss, learning capacity and academic performance</p></title>
  <aug>
    <au><snm>Curcio</snm><fnm>G</fnm></au>
    <au><snm>Ferrara</snm><fnm>M</fnm></au>
    <au><snm>De Gennaro</snm><fnm>L</fnm></au>
  </aug>
  <source>Sleep medicine reviews</source>
  <publisher>Elsevier</publisher>
  <pubdate>2006</pubdate>
  <volume>10</volume>
  <issue>5</issue>
  <fpage>323</fpage>
  <lpage>-337</lpage>
</bibl>

<bibl id="B5">
  <title><p>Physical activity and performance at school: a systematic review of
  the literature including a methodological quality assessment</p></title>
  <aug>
    <au><snm>Singh</snm><fnm>A</fnm></au>
    <au><snm>Uijtdewilligen</snm><fnm>L</fnm></au>
    <au><snm>Twisk</snm><fnm>JW</fnm></au>
    <au><snm>Van Mechelen</snm><fnm>W</fnm></au>
    <au><snm>Chinapaw</snm><fnm>MJ</fnm></au>
  </aug>
  <source>Archives of pediatrics \& adolescent medicine</source>
  <publisher>American Medical Association</publisher>
  <pubdate>2012</pubdate>
  <volume>166</volume>
  <issue>1</issue>
  <fpage>49</fpage>
  <lpage>-55</lpage>
</bibl>

<bibl id="B6">
  <title><p>Faking it: social desirability response bias in self-report
  research</p></title>
  <aug>
    <au><snm>Mortel</snm><fnm>TF</fnm></au>
    <au><cnm>others</cnm></au>
  </aug>
  <source>Australian Journal of Advanced Nursing, The</source>
  <publisher>Australian Nursing Federation</publisher>
  <pubdate>2008</pubdate>
  <volume>25</volume>
  <issue>4</issue>
  <fpage>40</fpage>
</bibl>

<bibl id="B7">
  <title><p>Comparing actual and self-reported measures of Facebook
  use</p></title>
  <aug>
    <au><snm>Junco</snm><fnm>R</fnm></au>
  </aug>
  <source>Computers in Human Behavior</source>
  <publisher>Elsevier</publisher>
  <pubdate>2013</pubdate>
  <volume>29</volume>
  <issue>3</issue>
  <fpage>626</fpage>
  <lpage>-631</lpage>
</bibl>

<bibl id="B8">
  <title><p>Systematic biases in social perception</p></title>
  <aug>
    <au><snm>Kumbasar</snm><fnm>E</fnm></au>
    <au><snm>Rommey</snm><fnm>AK</fnm></au>
    <au><snm>Batchelder</snm><fnm>WH</fnm></au>
  </aug>
  <source>American journal of sociology</source>
  <publisher>University of Chicago Press</publisher>
  <pubdate>1994</pubdate>
  <volume>100</volume>
  <issue>2</issue>
  <fpage>477</fpage>
  <lpage>-505</lpage>
</bibl>

<bibl id="B9">
  <title><p>How social exclusion distorts social network
  perceptions</p></title>
  <aug>
    <au><snm>O’Connor</snm><fnm>KM</fnm></au>
    <au><snm>Gladstone</snm><fnm>E</fnm></au>
  </aug>
  <source>Social Networks</source>
  <publisher>Elsevier</publisher>
  <pubdate>2015</pubdate>
  <volume>40</volume>
  <fpage>123</fpage>
  <lpage>-128</lpage>
</bibl>

<bibl id="B10">
  <title><p>Filling in the blanks: A theory of cognitive categories and the
  structure of social affiliation</p></title>
  <aug>
    <au><snm>Freeman</snm><fnm>LC</fnm></au>
  </aug>
  <source>Social Psychology Quarterly</source>
  <publisher>JSTOR</publisher>
  <pubdate>1992</pubdate>
  <fpage>118</fpage>
  <lpage>-127</lpage>
</bibl>

<bibl id="B11">
  <title><p>The problem of informant accuracy: The validity of retrospective
  data</p></title>
  <aug>
    <au><snm>Bernard</snm><fnm>HR</fnm></au>
    <au><snm>Killworth</snm><fnm>P</fnm></au>
    <au><snm>Kronenfeld</snm><fnm>D</fnm></au>
    <au><snm>Sailer</snm><fnm>L</fnm></au>
  </aug>
  <source>Annual review of anthropology</source>
  <publisher>Annual Reviews 4139 El Camino Way, PO Box 10139, Palo Alto, CA
  94303-0139, USA</publisher>
  <pubdate>1984</pubdate>
  <volume>13</volume>
  <issue>1</issue>
  <fpage>495</fpage>
  <lpage>-517</lpage>
</bibl>

<bibl id="B12">
  <title><p>Measuring large-scale social networks with high
  resolution</p></title>
  <aug>
    <au><snm>Stopczynski</snm><fnm>A</fnm></au>
    <au><snm>Sekara</snm><fnm>V</fnm></au>
    <au><snm>Sapiezynski</snm><fnm>P</fnm></au>
    <au><snm>Cuttone</snm><fnm>A</fnm></au>
    <au><snm>Madsen</snm><fnm>MM</fnm></au>
    <au><snm>Larsen</snm><fnm>JE</fnm></au>
    <au><snm>Lehmann</snm><fnm>S</fnm></au>
  </aug>
  <source>PloS one</source>
  <publisher>Public Library of Science</publisher>
  <pubdate>2014</pubdate>
  <volume>9</volume>
  <issue>4</issue>
  <fpage>e95978</fpage>
</bibl>

<bibl id="B13">
  <title><p>Predicting academic performance by data mining methods</p></title>
  <aug>
    <au><snm>Vandamme</snm><fnm>J P</fnm></au>
    <au><snm>Meskens</snm><fnm>N</fnm></au>
    <au><snm>Superby</snm><fnm>J F</fnm></au>
  </aug>
  <source>Education Economics</source>
  <publisher>Taylor \& Francis</publisher>
  <pubdate>2007</pubdate>
  <volume>15</volume>
  <issue>4</issue>
  <fpage>405</fpage>
  <lpage>-419</lpage>
</bibl>

<bibl id="B14">
  <title><p>Studentlife: assessing mental health, academic performance and
  behavioral trends of college students using smartphones</p></title>
  <aug>
    <au><snm>Wang</snm><fnm>R</fnm></au>
    <au><snm>Chen</snm><fnm>F</fnm></au>
    <au><snm>Chen</snm><fnm>Z</fnm></au>
    <au><snm>Li</snm><fnm>T</fnm></au>
    <au><snm>Harari</snm><fnm>G</fnm></au>
    <au><snm>Tignor</snm><fnm>S</fnm></au>
    <au><snm>Zhou</snm><fnm>X</fnm></au>
    <au><snm>Ben Zeev</snm><fnm>D</fnm></au>
    <au><snm>Campbell</snm><fnm>AT</fnm></au>
  </aug>
  <source>Proceedings of the 2014 ACM International Joint Conference on
  Pervasive and Ubiquitous Computing</source>
  <pubdate>2014</pubdate>
  <fpage>3</fpage>
  <lpage>-14</lpage>
</bibl>

<bibl id="B15">
  <title><p>A comparative analysis of techniques for predicting academic
  performance</p></title>
  <aug>
    <au><snm>Nghe</snm><fnm>NT</fnm></au>
    <au><snm>Janecek</snm><fnm>P</fnm></au>
    <au><snm>Haddawy</snm><fnm>P</fnm></au>
  </aug>
  <source>Frontiers In Education Conference-Global Engineering: Knowledge
  Without Borders, Opportunities Without Passports, 2007. FIE'07. 37th
  Annual</source>
  <pubdate>2007</pubdate>
  <fpage>T2G</fpage>
  <lpage>-7</lpage>
</bibl>

<bibl id="B16">
  <title><p>Relationship of initial class attendance and seating location to
  academic performance in psychology classes</p></title>
  <aug>
    <au><snm>Buckalew</snm><fnm>LW</fnm></au>
    <au><snm>Daly</snm><fnm>JD</fnm></au>
    <au><snm>Coffield</snm><fnm>KE</fnm></au>
  </aug>
  <source>Bulletin of the Psychonomic Society</source>
  <publisher>Springer</publisher>
  <pubdate>1986</pubdate>
  <volume>24</volume>
  <issue>1</issue>
  <fpage>63</fpage>
  <lpage>-64</lpage>
</bibl>

<bibl id="B17">
  <title><p>Does mandatory attendance improve student performance?</p></title>
  <aug>
    <au><snm>Marburger</snm><fnm>DR</fnm></au>
  </aug>
  <source>The Journal of Economic Education</source>
  <publisher>Taylor \& Francis</publisher>
  <pubdate>2006</pubdate>
  <volume>37</volume>
  <issue>2</issue>
  <fpage>148</fpage>
  <lpage>-155</lpage>
</bibl>

<bibl id="B18">
  <title><p>Class attendance and exam performance: A randomized
  experiment</p></title>
  <aug>
    <au><snm>Chen</snm><fnm>J</fnm></au>
    <au><snm>Lin</snm><fnm>TF</fnm></au>
  </aug>
  <source>The Journal of Economic Education</source>
  <publisher>Taylor \& Francis</publisher>
  <pubdate>2008</pubdate>
  <volume>39</volume>
  <issue>3</issue>
  <fpage>213</fpage>
  <lpage>-227</lpage>
</bibl>

<bibl id="B19">
  <title><p>Class attendance in college a meta-analytic review of the
  relationship of class attendance with grades and student
  characteristics</p></title>
  <aug>
    <au><snm>Cred{\'e}</snm><fnm>M</fnm></au>
    <au><snm>Roch</snm><fnm>SG</fnm></au>
    <au><snm>Kieszczynka</snm><fnm>UM</fnm></au>
  </aug>
  <source>Review of Educational Research</source>
  <publisher>SAGE Publications</publisher>
  <pubdate>2010</pubdate>
  <volume>80</volume>
  <issue>2</issue>
  <fpage>272</fpage>
  <lpage>-295</lpage>
</bibl>

<bibl id="B20">
  <title><p>Class attendance, peer similarity, and academic performance in a
  large field study</p></title>
  <aug>
    <au><snm>Kassarnig</snm><fnm>V</fnm></au>
    <au><snm>Bjerre Nielsen</snm><fnm>A</fnm></au>
    <au><snm>Mones</snm><fnm>E</fnm></au>
    <au><snm>Lehmann</snm><fnm>S</fnm></au>
    <au><snm>Lassen</snm><fnm>DD</fnm></au>
  </aug>
  <source>PLOS ONE</source>
  <publisher>Public Library of Science</publisher>
  <pubdate>2017</pubdate>
  <volume>12</volume>
  <issue>11</issue>
  <fpage>1</fpage>
  <lpage>15</lpage>
  <url>https://doi.org/10.1371/journal.pone.0187078</url>
</bibl>

<bibl id="B21">
  <title><p>Orderness Predicts Academic Performance: Behavioral Analysis on
  Campus Lifestyle</p></title>
  <aug>
    <au><snm>Cao</snm><fnm>Y</fnm></au>
    <au><snm>Lian</snm><fnm>D</fnm></au>
    <au><snm>Rong</snm><fnm>Z</fnm></au>
    <au><snm>Shi</snm><fnm>J</fnm></au>
    <au><snm>Wang</snm><fnm>Q</fnm></au>
    <au><snm>Wu</snm><fnm>Y</fnm></au>
    <au><snm>Yao</snm><fnm>H</fnm></au>
    <au><snm>Zhou</snm><fnm>T</fnm></au>
  </aug>
  <source>arXiv preprint arXiv:1704.04103</source>
  <pubdate>2017</pubdate>
</bibl>

<bibl id="B22">
  <title><p>Locus of control, study habits and attitudes, and college academic
  performance</p></title>
  <aug>
    <au><snm>Prociuk</snm><fnm>TJ</fnm></au>
    <au><snm>Breen</snm><fnm>LJ</fnm></au>
  </aug>
  <source>The Journal of Psychology</source>
  <publisher>Taylor \& Francis</publisher>
  <pubdate>1974</pubdate>
  <volume>88</volume>
  <issue>1</issue>
  <fpage>91</fpage>
  <lpage>-95</lpage>
</bibl>

<bibl id="B23">
  <title><p>The structure of phenotypic personality traits.</p></title>
  <aug>
    <au><snm>Goldberg</snm><fnm>LR</fnm></au>
  </aug>
  <source>American psychologist</source>
  <publisher>American Psychological Association</publisher>
  <pubdate>1993</pubdate>
  <volume>48</volume>
  <issue>1</issue>
  <fpage>26</fpage>
</bibl>

<bibl id="B24">
  <title><p>Personality and performance in “personality”: Conscientiousness
  and openness</p></title>
  <aug>
    <au><snm>Dollinger</snm><fnm>SJ</fnm></au>
    <au><snm>Orf</snm><fnm>LA</fnm></au>
  </aug>
  <source>Journal of Research in Personality</source>
  <publisher>Elsevier</publisher>
  <pubdate>1991</pubdate>
  <volume>25</volume>
  <issue>3</issue>
  <fpage>276</fpage>
  <lpage>-284</lpage>
</bibl>

<bibl id="B25">
  <title><p>Personality-intelligence relations: Assessment of typical
  intellectual engagement.</p></title>
  <aug>
    <au><snm>Goff</snm><fnm>M</fnm></au>
    <au><snm>Ackerman</snm><fnm>PL</fnm></au>
  </aug>
  <source>Journal of Educational Psychology</source>
  <publisher>American Psychological Association</publisher>
  <pubdate>1992</pubdate>
  <volume>84</volume>
  <issue>4</issue>
  <fpage>537</fpage>
</bibl>

<bibl id="B26">
  <title><p>Personality and cognitive ability predictors of performance in
  graduate business school.</p></title>
  <aug>
    <au><snm>Rothstein</snm><fnm>MG</fnm></au>
    <au><snm>Paunonen</snm><fnm>SV</fnm></au>
    <au><snm>Rush</snm><fnm>JC</fnm></au>
    <au><snm>King</snm><fnm>GA</fnm></au>
  </aug>
  <source>Journal of educational psychology</source>
  <publisher>American Psychological Association</publisher>
  <pubdate>1994</pubdate>
  <volume>86</volume>
  <issue>4</issue>
  <fpage>516</fpage>
</bibl>

<bibl id="B27">
  <title><p>Personality as a predictor of college performance</p></title>
  <aug>
    <au><snm>Wolfe</snm><fnm>RN</fnm></au>
    <au><snm>Johnson</snm><fnm>SD</fnm></au>
  </aug>
  <source>Educational and psychological measurement</source>
  <publisher>Sage Publications</publisher>
  <pubdate>1995</pubdate>
  <volume>55</volume>
  <issue>2</issue>
  <fpage>177</fpage>
  <lpage>-185</lpage>
</bibl>

<bibl id="B28">
  <title><p>Personality and interests as predictors of educational streaming
  and achievement</p></title>
  <aug>
    <au><snm>De Fruyt</snm><fnm>F</fnm></au>
    <au><snm>Mervielde</snm><fnm>I</fnm></au>
  </aug>
  <source>European journal of personality</source>
  <publisher>Wiley Online Library</publisher>
  <pubdate>1996</pubdate>
  <volume>10</volume>
  <issue>5</issue>
  <fpage>405</fpage>
  <lpage>-425</lpage>
</bibl>

<bibl id="B29">
  <title><p>Hierarchical organization of personality and prediction of
  behavior.</p></title>
  <aug>
    <au><snm>Paunonen</snm><fnm>SV</fnm></au>
  </aug>
  <source>Journal of Personality and Social Psychology</source>
  <publisher>American Psychological Association</publisher>
  <pubdate>1998</pubdate>
  <volume>74</volume>
  <issue>2</issue>
  <fpage>538</fpage>
</bibl>

<bibl id="B30">
  <title><p>Intellectual ability, learning style, personality, achievement
  motivation and academic success of psychology students in higher
  education</p></title>
  <aug>
    <au><snm>Busato</snm><fnm>VV</fnm></au>
    <au><snm>Prins</snm><fnm>FJ</fnm></au>
    <au><snm>Elshout</snm><fnm>JJ</fnm></au>
    <au><snm>Hamaker</snm><fnm>C</fnm></au>
  </aug>
  <source>Personality and Individual differences</source>
  <publisher>Elsevier</publisher>
  <pubdate>2000</pubdate>
  <volume>29</volume>
  <issue>6</issue>
  <fpage>1057</fpage>
  <lpage>-1068</lpage>
</bibl>

<bibl id="B31">
  <title><p>Big Five predictors of academic achievement</p></title>
  <aug>
    <au><snm>Paunonen</snm><fnm>SV</fnm></au>
    <au><snm>Ashton</snm><fnm>MC</fnm></au>
  </aug>
  <source>Journal of Research in Personality</source>
  <publisher>Elsevier</publisher>
  <pubdate>2001</pubdate>
  <volume>35</volume>
  <issue>1</issue>
  <fpage>78</fpage>
  <lpage>-90</lpage>
</bibl>

<bibl id="B32">
  <title><p>General and specific traits of personality and their relation to
  sleep and academic performance</p></title>
  <aug>
    <au><snm>Gray</snm><fnm>EK</fnm></au>
    <au><snm>Watson</snm><fnm>D</fnm></au>
  </aug>
  <source>Journal of personality</source>
  <publisher>Wiley Online Library</publisher>
  <pubdate>2002</pubdate>
  <volume>70</volume>
  <issue>2</issue>
  <fpage>177</fpage>
  <lpage>-206</lpage>
</bibl>

<bibl id="B33">
  <title><p>Medical students' personality characteristics and academic
  performance: A five-factor model perspective</p></title>
  <aug>
    <au><snm>Lievens</snm><fnm>F</fnm></au>
    <au><snm>Coetsier</snm><fnm>P</fnm></au>
    <au><snm>De Fruyt</snm><fnm>F</fnm></au>
    <au><snm>De Maeseneer</snm><fnm>J</fnm></au>
  </aug>
  <source>Medical education</source>
  <publisher>Wiley Online Library</publisher>
  <pubdate>2002</pubdate>
  <volume>36</volume>
  <issue>11</issue>
  <fpage>1050</fpage>
  <lpage>-1056</lpage>
</bibl>

<bibl id="B34">
  <title><p>The Effect of Personality and Precollege Characteristics on
  First-Year Activities and Academic Performance.</p></title>
  <aug>
    <au><snm>Bauer</snm><fnm>KW</fnm></au>
    <au><snm>Liang</snm><fnm>Q</fnm></au>
  </aug>
  <source>Journal of College Student Development</source>
  <publisher>ACPA Executive Office</publisher>
  <pubdate>2003</pubdate>
</bibl>

<bibl id="B35">
  <title><p>Personality traits and academic examination performance</p></title>
  <aug>
    <au><snm>Chamorro Premuzic</snm><fnm>T</fnm></au>
    <au><snm>Furnham</snm><fnm>A</fnm></au>
  </aug>
  <source>European Journal of Personality</source>
  <publisher>Wiley Online Library</publisher>
  <pubdate>2003</pubdate>
  <volume>17</volume>
  <issue>3</issue>
  <fpage>237</fpage>
  <lpage>-250</lpage>
</bibl>

<bibl id="B36">
  <title><p>Personality predicts academic performance: Evidence from two
  longitudinal university samples</p></title>
  <aug>
    <au><snm>Chamorro Premuzic</snm><fnm>T</fnm></au>
    <au><snm>Furnham</snm><fnm>A</fnm></au>
  </aug>
  <source>Journal of Research in Personality</source>
  <publisher>Elsevier</publisher>
  <pubdate>2003</pubdate>
  <volume>37</volume>
  <issue>4</issue>
  <fpage>319</fpage>
  <lpage>-338</lpage>
</bibl>

<bibl id="B37">
  <title><p>Personality and approaches to learning as predictors of academic
  achievement</p></title>
  <aug>
    <au><snm>Diseth</snm><fnm>{\AA}</fnm></au>
  </aug>
  <source>European Journal of personality</source>
  <publisher>Wiley Online Library</publisher>
  <pubdate>2003</pubdate>
  <volume>17</volume>
  <issue>2</issue>
  <fpage>143</fpage>
  <lpage>-155</lpage>
</bibl>

<bibl id="B38">
  <title><p>Individual differences and undergraduate academic success: The
  roles of personality, intelligence, and application</p></title>
  <aug>
    <au><snm>Farsides</snm><fnm>T</fnm></au>
    <au><snm>Woodfield</snm><fnm>R</fnm></au>
  </aug>
  <source>Personality and Individual differences</source>
  <publisher>Elsevier</publisher>
  <pubdate>2003</pubdate>
  <volume>34</volume>
  <issue>7</issue>
  <fpage>1225</fpage>
  <lpage>-1243</lpage>
</bibl>

<bibl id="B39">
  <title><p>Personality, cognitive ability, and beliefs about intelligence as
  predictors of academic performance</p></title>
  <aug>
    <au><snm>Furnham</snm><fnm>A</fnm></au>
    <au><snm>Chamorro Premuzic</snm><fnm>T</fnm></au>
    <au><snm>McDougall</snm><fnm>F</fnm></au>
  </aug>
  <source>Learning and Individual Differences</source>
  <publisher>Elsevier</publisher>
  <pubdate>2002</pubdate>
  <volume>14</volume>
  <issue>1</issue>
  <fpage>47</fpage>
  <lpage>-64</lpage>
</bibl>

<bibl id="B40">
  <title><p>Intelligence,“Big Five” personality traits, and work drive as
  predictors of course grade</p></title>
  <aug>
    <au><snm>Lounsbury</snm><fnm>JW</fnm></au>
    <au><snm>Sundstrom</snm><fnm>E</fnm></au>
    <au><snm>Loveland</snm><fnm>JM</fnm></au>
    <au><snm>Gibson</snm><fnm>LW</fnm></au>
  </aug>
  <source>Personality and Individual Differences</source>
  <publisher>Elsevier</publisher>
  <pubdate>2003</pubdate>
  <volume>35</volume>
  <issue>6</issue>
  <fpage>1231</fpage>
  <lpage>-1239</lpage>
</bibl>

<bibl id="B41">
  <title><p>Personality, cognition, and university students' examination
  performance</p></title>
  <aug>
    <au><snm>Phillips</snm><fnm>P</fnm></au>
    <au><snm>Abraham</snm><fnm>C</fnm></au>
    <au><snm>Bond</snm><fnm>R</fnm></au>
  </aug>
  <source>European Journal of Personality</source>
  <publisher>Wiley Online Library</publisher>
  <pubdate>2003</pubdate>
  <volume>17</volume>
  <issue>6</issue>
  <fpage>435</fpage>
  <lpage>-448</lpage>
</bibl>

<bibl id="B42">
  <title><p>The relationship between personality, approach to learning and
  academic performance</p></title>
  <aug>
    <au><snm>Duff</snm><fnm>A</fnm></au>
    <au><snm>Boyle</snm><fnm>E</fnm></au>
    <au><snm>Dunleavy</snm><fnm>K</fnm></au>
    <au><snm>Ferguson</snm><fnm>J</fnm></au>
  </aug>
  <source>Personality and individual differences</source>
  <publisher>Elsevier</publisher>
  <pubdate>2004</pubdate>
  <volume>36</volume>
  <issue>8</issue>
  <fpage>1907</fpage>
  <lpage>-1920</lpage>
</bibl>

<bibl id="B43">
  <title><p>Personality and intelligence as predictors of statistics
  examination grades</p></title>
  <aug>
    <au><snm>Furnham</snm><fnm>A</fnm></au>
    <au><snm>Chamorro Premuzic</snm><fnm>T</fnm></au>
  </aug>
  <source>Personality and Individual Differences</source>
  <publisher>Elsevier</publisher>
  <pubdate>2004</pubdate>
  <volume>37</volume>
  <issue>5</issue>
  <fpage>943</fpage>
  <lpage>-955</lpage>
</bibl>

<bibl id="B44">
  <title><p>The role of impulsivity in predicting maladaptive behaviour among
  female students</p></title>
  <aug>
    <au><snm>Hair</snm><fnm>P</fnm></au>
    <au><snm>Hampson</snm><fnm>SE</fnm></au>
  </aug>
  <source>Personality and Individual Differences</source>
  <publisher>Elsevier</publisher>
  <pubdate>2006</pubdate>
  <volume>40</volume>
  <issue>5</issue>
  <fpage>943</fpage>
  <lpage>-952</lpage>
</bibl>

<bibl id="B45">
  <title><p>Aptitude is not enough: How personality and behavior predict
  academic performance</p></title>
  <aug>
    <au><snm>Conard</snm><fnm>MA</fnm></au>
  </aug>
  <source>Journal of Research in Personality</source>
  <publisher>Elsevier</publisher>
  <pubdate>2006</pubdate>
  <volume>40</volume>
  <issue>3</issue>
  <fpage>339</fpage>
  <lpage>-346</lpage>
</bibl>

<bibl id="B46">
  <title><p>Does emotional intelligence assist in the prediction of academic
  success?</p></title>
  <aug>
    <au><snm>Barchard</snm><fnm>KA</fnm></au>
  </aug>
  <source>Educational and psychological measurement</source>
  <publisher>Sage Publications</publisher>
  <pubdate>2003</pubdate>
  <volume>63</volume>
  <issue>5</issue>
  <fpage>840</fpage>
  <lpage>-858</lpage>
</bibl>

<bibl id="B47">
  <title><p>A one-minute measure of the Big Five? Evaluating and abridging
  Shafer's (1999a) Big Five markers</p></title>
  <aug>
    <au><snm>Langford</snm><fnm>PH</fnm></au>
  </aug>
  <source>Personality and Individual Differences</source>
  <publisher>Elsevier</publisher>
  <pubdate>2003</pubdate>
  <volume>35</volume>
  <issue>5</issue>
  <fpage>1127</fpage>
  <lpage>-1140</lpage>
</bibl>

<bibl id="B48">
  <title><p>Developing a biodata measure and situational judgment inventory as
  predictors of college student performance.</p></title>
  <aug>
    <au><snm>Oswald</snm><fnm>FL</fnm></au>
    <au><snm>Schmitt</snm><fnm>N</fnm></au>
    <au><snm>Kim</snm><fnm>BH</fnm></au>
    <au><snm>Ramsay</snm><fnm>LJ</fnm></au>
    <au><snm>Gillespie</snm><fnm>MA</fnm></au>
  </aug>
  <source>Journal of applied psychology</source>
  <publisher>American Psychological Association</publisher>
  <pubdate>2004</pubdate>
  <volume>89</volume>
  <issue>2</issue>
  <fpage>187</fpage>
</bibl>

<bibl id="B49">
  <title><p>Sense of identity and collegiate academic achievement</p></title>
  <aug>
    <au><snm>Leong</snm><fnm>FT</fnm></au>
    <au><snm>Gibson</snm><fnm>LW</fnm></au>
    <au><snm>Lounsbury</snm><fnm>JW</fnm></au>
    <au><snm>Huffstetler</snm><fnm>BC</fnm></au>
  </aug>
  <source>Journal of College Student Development</source>
  <publisher>The Johns Hopkins University Press</publisher>
  <pubdate>2005</pubdate>
  <volume>46</volume>
  <issue>5</issue>
  <fpage>501</fpage>
  <lpage>-514</lpage>
</bibl>

<bibl id="B50">
  <title><p>Predicting academic success: General intelligence," Big Five"
  personality traits, and work drive</p></title>
  <aug>
    <au><snm>Ridgell</snm><fnm>SD</fnm></au>
    <au><snm>Lounsbury</snm><fnm>JW</fnm></au>
  </aug>
  <source>College Student Journal</source>
  <publisher>Project Innovation, Inc.</publisher>
  <pubdate>2004</pubdate>
  <volume>38</volume>
  <issue>4</issue>
  <fpage>607</fpage>
</bibl>

<bibl id="B51">
  <title><p>Role of the Big Five personality traits in predicting college
  students' academic motivation and achievement</p></title>
  <aug>
    <au><snm>Komarraju</snm><fnm>M</fnm></au>
    <au><snm>Karau</snm><fnm>SJ</fnm></au>
    <au><snm>Schmeck</snm><fnm>RR</fnm></au>
  </aug>
  <source>Learning and individual differences</source>
  <publisher>Elsevier</publisher>
  <pubdate>2009</pubdate>
  <volume>19</volume>
  <issue>1</issue>
  <fpage>47</fpage>
  <lpage>-52</lpage>
</bibl>

<bibl id="B52">
  <title><p>Personality predictors of academic outcomes: big five correlates of
  GPA and SAT scores.</p></title>
  <aug>
    <au><snm>Noftle</snm><fnm>EE</fnm></au>
    <au><snm>Robins</snm><fnm>RW</fnm></au>
  </aug>
  <source>Journal of personality and social psychology</source>
  <publisher>American Psychological Association</publisher>
  <pubdate>2007</pubdate>
  <volume>93</volume>
  <issue>1</issue>
  <fpage>116</fpage>
</bibl>

<bibl id="B53">
  <title><p>Self-efficacy, self-esteem and their impact on academic
  performance</p></title>
  <aug>
    <au><snm>Lane</snm><fnm>J</fnm></au>
    <au><snm>Lane</snm><fnm>AM</fnm></au>
    <au><snm>Kyprianou</snm><fnm>A</fnm></au>
  </aug>
  <source>Social Behavior and Personality: an international journal</source>
  <publisher>Scientific Journal Publishers</publisher>
  <pubdate>2004</pubdate>
  <volume>32</volume>
  <issue>3</issue>
  <fpage>247</fpage>
  <lpage>-256</lpage>
</bibl>

<bibl id="B54">
  <title><p>The relationship between cell phone use, academic performance,
  anxiety, and satisfaction with life in college students</p></title>
  <aug>
    <au><snm>Lepp</snm><fnm>A</fnm></au>
    <au><snm>Barkley</snm><fnm>JE</fnm></au>
    <au><snm>Karpinski</snm><fnm>AC</fnm></au>
  </aug>
  <source>Computers in Human Behavior</source>
  <publisher>Elsevier</publisher>
  <pubdate>2014</pubdate>
  <volume>31</volume>
  <fpage>343</fpage>
  <lpage>-350</lpage>
</bibl>

<bibl id="B55">
  <title><p>Life satisfaction among university students in a Canadian prairie
  city: A multivariate analysis</p></title>
  <aug>
    <au><snm>Chow</snm><fnm>HP</fnm></au>
  </aug>
  <source>Social Indicators Research</source>
  <publisher>Springer</publisher>
  <pubdate>2005</pubdate>
  <volume>70</volume>
  <issue>2</issue>
  <fpage>139</fpage>
  <lpage>-150</lpage>
</bibl>

<bibl id="B56">
  <title><p>Relationships of personality, affect, emotional intelligence and
  coping with student stress and academic success: Different patterns of
  association for stress and success</p></title>
  <aug>
    <au><snm>Saklofske</snm><fnm>DH</fnm></au>
    <au><snm>Austin</snm><fnm>EJ</fnm></au>
    <au><snm>Mastoras</snm><fnm>SM</fnm></au>
    <au><snm>Beaton</snm><fnm>L</fnm></au>
    <au><snm>Osborne</snm><fnm>SE</fnm></au>
  </aug>
  <source>Learning and Individual Differences</source>
  <publisher>Elsevier</publisher>
  <pubdate>2012</pubdate>
  <volume>22</volume>
  <issue>2</issue>
  <fpage>251</fpage>
  <lpage>-257</lpage>
</bibl>

<bibl id="B57">
  <title><p>A prospective analysis of stress and academic performance in the
  first two years of medical school</p></title>
  <aug>
    <au><snm>Stewart</snm><fnm>SM</fnm></au>
    <au><snm>Lam</snm><fnm>TH</fnm></au>
    <au><snm>Betson</snm><fnm>CL</fnm></au>
    <au><snm>Wong</snm><fnm>CM</fnm></au>
    <au><snm>Wong</snm><fnm>AMP</fnm></au>
  </aug>
  <source>Medical Education-Oxford</source>
  <publisher>Wiley Online Library</publisher>
  <pubdate>1999</pubdate>
  <volume>33</volume>
  <issue>4</issue>
  <fpage>243</fpage>
  <lpage>-250</lpage>
</bibl>

<bibl id="B58">
  <title><p>Learned resourcefulness moderates the relationship between academic
  stress and academic performance</p></title>
  <aug>
    <au><snm>Akgun</snm><fnm>S</fnm></au>
    <au><snm>Ciarrochi</snm><fnm>J</fnm></au>
  </aug>
  <source>Educational Psychology</source>
  <publisher>Taylor \& Francis</publisher>
  <pubdate>2003</pubdate>
  <volume>23</volume>
  <issue>3</issue>
  <fpage>287</fpage>
  <lpage>-294</lpage>
</bibl>

<bibl id="B59">
  <title><p>The effects of depressed mood on academic performance in college
  students.</p></title>
  <aug>
    <au><snm>Haines</snm><fnm>ME</fnm></au>
    <au><snm>Norris</snm><fnm>MP</fnm></au>
    <au><snm>Kashy</snm><fnm>DA</fnm></au>
  </aug>
  <source>Journal of College Student Development</source>
  <publisher>ACPA Executive Office</publisher>
  <pubdate>1996</pubdate>
</bibl>

<bibl id="B60">
  <title><p>The relationship between depression and college academic
  performance</p></title>
  <aug>
    <au><snm>Leach</snm><fnm>J</fnm></au>
  </aug>
  <source>College Student Journal</source>
  <publisher>Project Innovation, Inc.</publisher>
  <pubdate>2009</pubdate>
  <volume>43</volume>
  <issue>2</issue>
  <fpage>325</fpage>
</bibl>

<bibl id="B61">
  <title><p>Anxiety and depression in academic performance: An exploration of
  the mediating factors of worry and working memory</p></title>
  <aug>
    <au><snm>Owens</snm><fnm>M</fnm></au>
    <au><snm>Stevenson</snm><fnm>J</fnm></au>
    <au><snm>Hadwin</snm><fnm>JA</fnm></au>
    <au><snm>Norgate</snm><fnm>R</fnm></au>
  </aug>
  <source>School Psychology International</source>
  <publisher>SAGE Publications</publisher>
  <pubdate>2012</pubdate>
  <volume>33</volume>
  <issue>4</issue>
  <fpage>433</fpage>
  <lpage>-449</lpage>
</bibl>

<bibl id="B62">
  <title><p>The Impact of Social Media Networks Websites Usage on Students’
  Academic Performance</p></title>
  <aug>
    <au><snm>Maqableh</snm><fnm>MM</fnm></au>
    <au><snm>Rajab</snm><fnm>L</fnm></au>
    <au><snm>Quteshat</snm><fnm>W</fnm></au>
    <au><snm>Moh’d Taisir Masa</snm><fnm>R</fnm></au>
    <au><snm>Khatib</snm><fnm>T</fnm></au>
    <au><snm>Karajeh</snm><fnm>H</fnm></au>
    <au><cnm>others</cnm></au>
  </aug>
  <source>Communications and Network</source>
  <publisher>Scientific Research Publishing</publisher>
  <pubdate>2015</pubdate>
  <volume>7</volume>
  <issue>04</issue>
  <fpage>159</fpage>
</bibl>

<bibl id="B63">
  <title><p>Social Media Use, Engagement and Addiction as Predictors of
  Academic Performance</p></title>
  <aug>
    <au><snm>Al Menayes</snm><fnm>JJ</fnm></au>
  </aug>
  <source>International Journal of Psychological Studies</source>
  <pubdate>2015</pubdate>
  <volume>7</volume>
  <issue>4</issue>
  <fpage>86</fpage>
</bibl>

<bibl id="B64">
  <title><p>The relationship between mobile social media use and academic
  performance in university students</p></title>
  <aug>
    <au><snm>Al Menayes</snm><fnm>JJ</fnm></au>
  </aug>
  <source>New Media and Mass Communication</source>
  <pubdate>2014</pubdate>
  <volume>25</volume>
  <fpage>23</fpage>
  <lpage>-29</lpage>
</bibl>

<bibl id="B65">
  <title><p>An exploration of social networking site use, multitasking, and
  academic performance among United States and European university
  students</p></title>
  <aug>
    <au><snm>Karpinski</snm><fnm>AC</fnm></au>
    <au><snm>Kirschner</snm><fnm>PA</fnm></au>
    <au><snm>Ozer</snm><fnm>I</fnm></au>
    <au><snm>Mellott</snm><fnm>JA</fnm></au>
    <au><snm>Ochwo</snm><fnm>P</fnm></au>
  </aug>
  <source>Computers in Human Behavior</source>
  <publisher>Elsevier</publisher>
  <pubdate>2013</pubdate>
  <volume>29</volume>
  <issue>3</issue>
  <fpage>1182</fpage>
  <lpage>-1192</lpage>
</bibl>

<bibl id="B66">
  <title><p>Effect of online social networking on student academic
  performance</p></title>
  <aug>
    <au><snm>Paul</snm><fnm>JA</fnm></au>
    <au><snm>Baker</snm><fnm>HM</fnm></au>
    <au><snm>Cochran</snm><fnm>JD</fnm></au>
  </aug>
  <source>Computers in Human Behavior</source>
  <publisher>Elsevier</publisher>
  <pubdate>2012</pubdate>
  <volume>28</volume>
  <issue>6</issue>
  <fpage>2117</fpage>
  <lpage>-2127</lpage>
</bibl>

<bibl id="B67">
  <title><p>The relationship between frequency of Facebook use, participation
  in Facebook activities, and student engagement</p></title>
  <aug>
    <au><snm>Junco</snm><fnm>R</fnm></au>
  </aug>
  <source>Computers \& Education</source>
  <publisher>Elsevier</publisher>
  <pubdate>2012</pubdate>
  <volume>58</volume>
  <issue>1</issue>
  <fpage>162</fpage>
  <lpage>-171</lpage>
</bibl>

<bibl id="B68">
  <title><p>The wired generation: Academic and social outcomes of electronic
  media use among university students</p></title>
  <aug>
    <au><snm>Jacobsen</snm><fnm>WC</fnm></au>
    <au><snm>Forste</snm><fnm>R</fnm></au>
  </aug>
  <source>Cyberpsychology, Behavior, and Social Networking</source>
  <publisher>Mary Ann Liebert, Inc. 140 Huguenot Street, 3rd Floor New
  Rochelle, NY 10801 USA</publisher>
  <pubdate>2011</pubdate>
  <volume>14</volume>
  <issue>5</issue>
  <fpage>275</fpage>
  <lpage>-280</lpage>
</bibl>

<bibl id="B69">
  <title><p>Facebook{\textregistered} and academic performance</p></title>
  <aug>
    <au><snm>Kirschner</snm><fnm>PA</fnm></au>
    <au><snm>Karpinski</snm><fnm>AC</fnm></au>
  </aug>
  <source>Computers in human behavior</source>
  <publisher>Elsevier</publisher>
  <pubdate>2010</pubdate>
  <volume>26</volume>
  <issue>6</issue>
  <fpage>1237</fpage>
  <lpage>-1245</lpage>
</bibl>

<bibl id="B70">
  <title><p>Facebook and academic performance: Reconciling a media sensation
  with data</p></title>
  <aug>
    <au><snm>Pasek</snm><fnm>J</fnm></au>
    <au><snm>Hargittai</snm><fnm>E</fnm></au>
    <au><cnm>others</cnm></au>
  </aug>
  <source>First Monday</source>
  <pubdate>2009</pubdate>
  <volume>14</volume>
  <issue>5</issue>
</bibl>

<bibl id="B71">
  <title><p>Facebook usage, socialization and academic performance</p></title>
  <aug>
    <au><snm>Ainin</snm><fnm>S</fnm></au>
    <au><snm>Naqshbandi</snm><fnm>MM</fnm></au>
    <au><snm>Moghavvemi</snm><fnm>S</fnm></au>
    <au><snm>Jaafar</snm><fnm>NI</fnm></au>
  </aug>
  <source>Computers \& Education</source>
  <publisher>Elsevier</publisher>
  <pubdate>2015</pubdate>
  <volume>83</volume>
  <fpage>64</fpage>
  <lpage>-73</lpage>
</bibl>

<bibl id="B72">
  <title><p>Online disclosure: An empirical examination of undergraduate
  Facebook profiles</p></title>
  <aug>
    <au><snm>Kolek</snm><fnm>EA</fnm></au>
    <au><snm>Saunders</snm><fnm>D</fnm></au>
  </aug>
  <source>NASPA journal</source>
  <publisher>Taylor \& Francis</publisher>
  <pubdate>2008</pubdate>
  <volume>45</volume>
  <issue>1</issue>
  <fpage>1</fpage>
  <lpage>-25</lpage>
</bibl>

<bibl id="B73">
  <title><p>Social Network: Academic &amp Social Impact on College
  Students</p></title>
  <aug>
    <au><snm>Tayseer</snm><fnm>M</fnm></au>
    <au><snm>Zoghieb</snm><fnm>F</fnm></au>
    <au><snm>Alcheikh</snm><fnm>I</fnm></au>
    <au><snm>Awadallah</snm><fnm>MN</fnm></au>
  </aug>
  <publisher>Retrieved 20th November</publisher>
  <pubdate>2014</pubdate>
</bibl>

<bibl id="B74">
  <title><p>Peer Effects with Random Assignment: Results for Dartmouth
  Roommates</p></title>
  <aug>
    <au><snm>Sacerdote</snm><fnm>B</fnm></au>
  </aug>
  <source>Quarterly Journal of Economics</source>
  <publisher>JSTOR</publisher>
  <pubdate>2001</pubdate>
  <fpage>681</fpage>
  <lpage>-704</lpage>
</bibl>

<bibl id="B75">
  <title><p>Peer effects in academic outcomes: Evidence from a natural
  experiment</p></title>
  <aug>
    <au><snm>Zimmerman</snm><fnm>DJ</fnm></au>
  </aug>
  <source>Review of Economics and statistics</source>
  <publisher>MIT Press</publisher>
  <pubdate>2003</pubdate>
  <volume>85</volume>
  <issue>1</issue>
  <fpage>9</fpage>
  <lpage>-23</lpage>
</bibl>

<bibl id="B76">
  <title><p>What can be learned about peer effects using college roommates?
  Evidence from new survey data and students from disadvantaged
  backgrounds</p></title>
  <aug>
    <au><snm>Stinebrickner</snm><fnm>R</fnm></au>
    <au><snm>Stinebrickner</snm><fnm>TR</fnm></au>
  </aug>
  <source>Journal of public Economics</source>
  <publisher>Elsevier</publisher>
  <pubdate>2006</pubdate>
  <volume>90</volume>
  <issue>8</issue>
  <fpage>1435</fpage>
  <lpage>-1454</lpage>
</bibl>

<bibl id="B77">
  <title><p>From natural variation to optimal policy? The importance of
  endogenous peer group formation</p></title>
  <aug>
    <au><snm>Carrell</snm><fnm>SE</fnm></au>
    <au><snm>Sacerdote</snm><fnm>BI</fnm></au>
    <au><snm>West</snm><fnm>JE</fnm></au>
  </aug>
  <source>Econometrica</source>
  <publisher>Wiley Online Library</publisher>
  <pubdate>2013</pubdate>
  <volume>81</volume>
  <issue>3</issue>
  <fpage>855</fpage>
  <lpage>-882</lpage>
</bibl>

<bibl id="B78">
  <title><p>Examining the effect of social influence on student performance
  through network autocorrelation models</p></title>
  <aug>
    <au><snm>Vitale</snm><fnm>MP</fnm></au>
    <au><snm>Porzio</snm><fnm>GC</fnm></au>
    <au><snm>Doreian</snm><fnm>P</fnm></au>
  </aug>
  <source>Journal of Applied Statistics</source>
  <publisher>Taylor \& Francis</publisher>
  <pubdate>2016</pubdate>
  <volume>43</volume>
  <issue>1</issue>
  <fpage>115</fpage>
  <lpage>-127</lpage>
</bibl>

<bibl id="B79">
  <title><p>Formation of homophily in academic performance: students prefer to
  change their friends rather than performance</p></title>
  <aug>
    <au><snm>Smirnov</snm><fnm>I</fnm></au>
    <au><snm>Thurner</snm><fnm>S</fnm></au>
  </aug>
  <source>arXiv preprint arXiv:1606.09082</source>
  <pubdate>2016</pubdate>
</bibl>

<bibl id="B80">
  <title><p>How social ties affect peer group effects: Case of university
  students</p></title>
  <aug>
    <au><snm>Poldin</snm><fnm>O</fnm></au>
    <au><snm>Valeeva</snm><fnm>D</fnm></au>
    <au><snm>Yudkevich</snm><fnm>M</fnm></au>
  </aug>
  <source>Higher School of Economics Research Paper No. WP BPR</source>
  <pubdate>2013</pubdate>
  <volume>15</volume>
</bibl>

<bibl id="B81">
  <title><p>The old boy (and girl) network: Social network formation on
  university campuses</p></title>
  <aug>
    <au><snm>Mayer</snm><fnm>A</fnm></au>
    <au><snm>Puller</snm><fnm>SL</fnm></au>
  </aug>
  <source>Journal of public economics</source>
  <publisher>Elsevier</publisher>
  <pubdate>2008</pubdate>
  <volume>92</volume>
  <issue>1</issue>
  <fpage>329</fpage>
  <lpage>-347</lpage>
</bibl>

<bibl id="B82">
  <title><p>Focused activities and the development of social capital in a
  distributed learning “community”</p></title>
  <aug>
    <au><snm>Yuan</snm><fnm>YC</fnm></au>
    <au><snm>Gay</snm><fnm>G</fnm></au>
    <au><snm>Hembrooke</snm><fnm>H</fnm></au>
  </aug>
  <source>The Information Society</source>
  <publisher>Taylor \& Francis</publisher>
  <pubdate>2006</pubdate>
  <volume>22</volume>
  <issue>1</issue>
  <fpage>25</fpage>
  <lpage>-39</lpage>
</bibl>

<bibl id="B83">
  <title><p>It’s not just what you know, it’s who you know: Testing a model
  of the relative importance of social networks to academic
  performance</p></title>
  <aug>
    <au><snm>Rizzuto</snm><fnm>TE</fnm></au>
    <au><snm>LeDoux</snm><fnm>J</fnm></au>
    <au><snm>Hatala</snm><fnm>JP</fnm></au>
  </aug>
  <source>Social Psychology of Education</source>
  <publisher>Springer</publisher>
  <pubdate>2009</pubdate>
  <volume>12</volume>
  <issue>2</issue>
  <fpage>175</fpage>
  <lpage>-189</lpage>
</bibl>

<bibl id="B84">
  <title><p>The influence of relationship networks on academic performance in
  higher education: a comparative study between students of a creative and a
  non-creative discipline</p></title>
  <aug>
    <au><snm>Tom{\'a}s Miquel</snm><fnm>JV</fnm></au>
    <au><snm>Exp{\'o}sito Langa</snm><fnm>M</fnm></au>
    <au><snm>Nicolau Juli{\'a}</snm><fnm>D</fnm></au>
  </aug>
  <source>Higher Education</source>
  <publisher>Springer</publisher>
  <pubdate>2015</pubdate>
  <fpage>1</fpage>
  <lpage>-16</lpage>
</bibl>

<bibl id="B85">
  <title><p>Social networks and the performance of individuals and
  groups</p></title>
  <aug>
    <au><snm>Sparrowe</snm><fnm>RT</fnm></au>
    <au><snm>Liden</snm><fnm>RC</fnm></au>
    <au><snm>Wayne</snm><fnm>SJ</fnm></au>
    <au><snm>Kraimer</snm><fnm>ML</fnm></au>
  </aug>
  <source>Academy of management journal</source>
  <publisher>Academy of Management</publisher>
  <pubdate>2001</pubdate>
  <volume>44</volume>
  <issue>2</issue>
  <fpage>316</fpage>
  <lpage>-325</lpage>
</bibl>

<bibl id="B86">
  <title><p>“Psst… What do you think?” The relationship between advice
  prestige, type of advice, and academic performance</p></title>
  <aug>
    <au><snm>Smith</snm><fnm>RA</fnm></au>
    <au><snm>Peterson</snm><fnm>BL</fnm></au>
  </aug>
  <source>Communication Education</source>
  <publisher>Taylor \& Francis</publisher>
  <pubdate>2007</pubdate>
  <volume>56</volume>
  <issue>3</issue>
  <fpage>278</fpage>
  <lpage>-291</lpage>
</bibl>

<bibl id="B87">
  <title><p>Visualising the invisible: a network approach to reveal the
  informal social side of student learning</p></title>
  <aug>
    <au><snm>Hommes</snm><fnm>J</fnm></au>
    <au><snm>Rienties</snm><fnm>B</fnm></au>
    <au><snm>De Grave</snm><fnm>W</fnm></au>
    <au><snm>Bos</snm><fnm>G</fnm></au>
    <au><snm>Schuwirth</snm><fnm>L</fnm></au>
    <au><snm>Scherpbier</snm><fnm>A</fnm></au>
  </aug>
  <source>Advances in Health Sciences Education</source>
  <publisher>Springer</publisher>
  <pubdate>2012</pubdate>
  <volume>17</volume>
  <issue>5</issue>
  <fpage>743</fpage>
  <lpage>-757</lpage>
</bibl>

<bibl id="B88">
  <title><p>The social fabric of a team-based MBA program: Network effects on
  student satisfaction and performance</p></title>
  <aug>
    <au><snm>Baldwin</snm><fnm>TT</fnm></au>
    <au><snm>Bedell</snm><fnm>MD</fnm></au>
    <au><snm>Johnson</snm><fnm>JL</fnm></au>
  </aug>
  <source>Academy of management journal</source>
  <publisher>Academy of Management</publisher>
  <pubdate>1997</pubdate>
  <volume>40</volume>
  <issue>6</issue>
  <fpage>1369</fpage>
  <lpage>-1397</lpage>
</bibl>

<bibl id="B89">
  <title><p>Effects of Social Network On Students, Performance: A web-based
  Forum Study in Taiwan. JALN, 7 (3), 93. Retrieved 11 26, 2011</p></title>
  <aug>
    <au><snm>Yang</snm><fnm>HL</fnm></au>
    <au><snm>Tang</snm><fnm>JH</fnm></au>
  </aug>
  <pubdate>2003</pubdate>
</bibl>

<bibl id="B90">
  <title><p>Social networks, communication styles, and learning performance in
  a CSCL community</p></title>
  <aug>
    <au><snm>Cho</snm><fnm>H</fnm></au>
    <au><snm>Gay</snm><fnm>G</fnm></au>
    <au><snm>Davidson</snm><fnm>B</fnm></au>
    <au><snm>Ingraffea</snm><fnm>A</fnm></au>
  </aug>
  <source>Computers \& Education</source>
  <publisher>Elsevier</publisher>
  <pubdate>2007</pubdate>
  <volume>49</volume>
  <issue>2</issue>
  <fpage>309</fpage>
  <lpage>-329</lpage>
</bibl>

<bibl id="B91">
  <title><p>Ties that bind: A social network approach to understanding student
  integration and persistence</p></title>
  <aug>
    <au><snm>Thomas</snm><fnm>SL</fnm></au>
  </aug>
  <source>Journal of Higher Education</source>
  <publisher>JSTOR</publisher>
  <pubdate>2000</pubdate>
  <fpage>591</fpage>
  <lpage>-615</lpage>
</bibl>

<bibl id="B92">
  <title><p>Structuring groups for cooperative learning</p></title>
  <aug>
    <au><snm>Johnson</snm><fnm>DW</fnm></au>
    <au><snm>Johnson</snm><fnm>RT</fnm></au>
  </aug>
  <source>Journal of Management Education</source>
  <publisher>Sage Publications</publisher>
  <pubdate>1984</pubdate>
  <volume>9</volume>
  <issue>4</issue>
  <fpage>8</fpage>
  <lpage>-17</lpage>
</bibl>

<bibl id="B93">
  <title><p>Identification of endogenous social effects: The reflection
  problem</p></title>
  <aug>
    <au><snm>Manski</snm><fnm>CF</fnm></au>
  </aug>
  <source>The review of economic studies</source>
  <publisher>Oxford University Press</publisher>
  <pubdate>1993</pubdate>
  <volume>60</volume>
  <issue>3</issue>
  <fpage>531</fpage>
  <lpage>-542</lpage>
</bibl>

<bibl id="B94">
  <title><p>Mobile phone data for inferring social network
  structure</p></title>
  <aug>
    <au><snm>Eagle</snm><fnm>N</fnm></au>
    <au><snm>Pentland</snm><fnm>AS</fnm></au>
    <au><snm>Lazer</snm><fnm>D</fnm></au>
  </aug>
  <source>Social computing, behavioral modeling, and prediction</source>
  <publisher>Springer</publisher>
  <pubdate>2008</pubdate>
  <fpage>79</fpage>
  <lpage>-88</lpage>
</bibl>

<bibl id="B95">
  <title><p>Measuring tie strength</p></title>
  <aug>
    <au><snm>Marsden</snm><fnm>PV</fnm></au>
    <au><snm>Campbell</snm><fnm>KE</fnm></au>
  </aug>
  <source>Social forces</source>
  <publisher>The University of North Carolina Press</publisher>
  <pubdate>1984</pubdate>
  <volume>63</volume>
  <issue>2</issue>
  <fpage>482</fpage>
  <lpage>-501</lpage>
</bibl>

<bibl id="B96">
  <title><p>Children's friendship relations: A meta-analytic
  review.</p></title>
  <aug>
    <au><snm>Newcomb</snm><fnm>AF</fnm></au>
    <au><snm>Bagwell</snm><fnm>CL</fnm></au>
  </aug>
  <source>Psychological bulletin</source>
  <publisher>American Psychological Association</publisher>
  <pubdate>1995</pubdate>
  <volume>117</volume>
  <issue>2</issue>
  <fpage>306</fpage>
</bibl>

<bibl id="B97">
  <title><p>The strength of friendship ties in proximity sensor
  data</p></title>
  <aug>
    <au><snm>Sekara</snm><fnm>V</fnm></au>
    <au><snm>Lehmann</snm><fnm>S</fnm></au>
  </aug>
  <source>PloS one</source>
  <publisher>Public Library of Science</publisher>
  <pubdate>2014</pubdate>
  <volume>9</volume>
  <issue>7</issue>
  <fpage>e100915</fpage>
</bibl>

<bibl id="B98">
  <title><p>Inferring person-to-person proximity using WiFi signals</p></title>
  <aug>
    <au><snm>Sapiezynski</snm><fnm>P</fnm></au>
    <au><snm>Stopczynski</snm><fnm>A</fnm></au>
    <au><snm>Wind</snm><fnm>DK</fnm></au>
    <au><snm>Leskovec</snm><fnm>J</fnm></au>
    <au><snm>Lehmann</snm><fnm>S</fnm></au>
  </aug>
  <source>Proceedings of the ACM on Interactive, Mobile, Wearable and
  Ubiquitous Technologies</source>
  <publisher>ACM</publisher>
  <pubdate>2017</pubdate>
  <volume>1</volume>
  <issue>2</issue>
  <fpage>24</fpage>
</bibl>

<bibl id="B99">
  <title><p>Inferring friendship network structure by using mobile phone
  data</p></title>
  <aug>
    <au><snm>Eagle</snm><fnm>N</fnm></au>
    <au><snm>Pentland</snm><fnm>AS</fnm></au>
    <au><snm>Lazer</snm><fnm>D</fnm></au>
  </aug>
  <source>Proceedings of the national academy of sciences</source>
  <publisher>National Acad Sciences</publisher>
  <pubdate>2009</pubdate>
  <volume>106</volume>
  <issue>36</issue>
  <fpage>15274</fpage>
  <lpage>-15278</lpage>
</bibl>

<bibl id="B100">
  <title><p>Applications of machine learning in animal behaviour
  studies</p></title>
  <aug>
    <au><snm>Valletta</snm><fnm>JJ</fnm></au>
    <au><snm>Torney</snm><fnm>C</fnm></au>
    <au><snm>Kings</snm><fnm>M</fnm></au>
    <au><snm>Thornton</snm><fnm>A</fnm></au>
    <au><snm>Madden</snm><fnm>J</fnm></au>
  </aug>
  <source>Animal Behaviour</source>
  <publisher>Elsevier</publisher>
  <pubdate>2017</pubdate>
  <volume>124</volume>
  <fpage>203</fpage>
  <lpage>-220</lpage>
</bibl>

<bibl id="B101">
  <title><p>“I’ll see you on IM, text, or call you”: A social network
  approach of adolescents’ use of communication media</p></title>
  <aug>
    <au><snm>Van Cleemput</snm><fnm>K</fnm></au>
  </aug>
  <source>Bulletin of Science, Technology \& Society</source>
  <publisher>Sage Publications</publisher>
  <pubdate>2010</pubdate>
  <volume>30</volume>
  <issue>2</issue>
  <fpage>75</fpage>
  <lpage>-85</lpage>
</bibl>

<bibl id="B102">
  <title><p>The effects of attendance on academic performance: Panel data
  evidence for introductory microeconomics</p></title>
  <aug>
    <au><snm>Stanca</snm><fnm>L</fnm></au>
  </aug>
  <source>The Journal of Economic Education</source>
  <publisher>Taylor \& Francis</publisher>
  <pubdate>2006</pubdate>
  <volume>37</volume>
  <issue>3</issue>
  <fpage>251</fpage>
  <lpage>-266</lpage>
</bibl>

<bibl id="B103">
  <title><p>Class attendance in undergraduate courses</p></title>
  <aug>
    <au><snm>Van Blerkom</snm><fnm>ML</fnm></au>
  </aug>
  <source>The Journal of psychology</source>
  <publisher>Taylor \& Francis</publisher>
  <pubdate>1992</pubdate>
  <volume>126</volume>
  <issue>5</issue>
  <fpage>487</fpage>
  <lpage>-494</lpage>
</bibl>

<bibl id="B104">
  <title><p>How much does coming to class matter? Some evidence of class
  attendance and grade performance.</p></title>
  <aug>
    <au><snm>Brocato</snm><fnm>J</fnm></au>
  </aug>
  <source>Educational Research Quarterly</source>
  <publisher>U. of Southern Calif. Sch. of Educ</publisher>
  <pubdate>1989</pubdate>
</bibl>

<bibl id="B105">
  <title><p>The cost of cutting class: Attendance as a predictor of
  success</p></title>
  <aug>
    <au><snm>Gump</snm><fnm>SE</fnm></au>
  </aug>
  <source>College Teaching</source>
  <publisher>Taylor \& Francis</publisher>
  <pubdate>2005</pubdate>
  <volume>53</volume>
  <issue>1</issue>
  <fpage>21</fpage>
  <lpage>-26</lpage>
</bibl>

<bibl id="B106">
  <title><p>Cumulative class attendance and exam performance</p></title>
  <aug>
    <au><snm>Lin</snm><fnm>TF</fnm></au>
    <au><snm>Chen</snm><fnm>J</fnm></au>
  </aug>
  <source>Applied Economics Letters</source>
  <publisher>Taylor \& Francis</publisher>
  <pubdate>2006</pubdate>
  <volume>13</volume>
  <issue>14</issue>
  <fpage>937</fpage>
  <lpage>-942</lpage>
</bibl>

<bibl id="B107">
  <title><p>Personality and assessment</p></title>
  <aug>
    <au><snm>Mischel</snm><fnm>W</fnm></au>
  </aug>
  <publisher>New Jersey, USA: Lawrence Erlbaum Associates</publisher>
  <pubdate>2013</pubdate>
</bibl>

<bibl id="B108">
  <title><p>Bias in observational studies</p></title>
  <aug>
    <au><snm>Hill</snm><fnm>HA</fnm></au>
    <au><snm>Kleinbaum</snm><fnm>DG</fnm></au>
  </aug>
  <source>Encyclopedia of Biostatistics</source>
  <publisher>Wiley Online Library</publisher>
  <pubdate>2000</pubdate>
</bibl>

<bibl id="B109">
  <title><p>Predictors of academic achievement and retention among college
  freshmen: A longitudinal study</p></title>
  <aug>
    <au><snm>DeBerard</snm><fnm>MS</fnm></au>
    <au><snm>Spielmans</snm><fnm>G</fnm></au>
    <au><snm>Julka</snm><fnm>D</fnm></au>
  </aug>
  <source>College student journal</source>
  <publisher>PROJECT INNOVATION INC</publisher>
  <pubdate>2004</pubdate>
  <volume>38</volume>
  <issue>1</issue>
  <fpage>66</fpage>
  <lpage>-80</lpage>
</bibl>

<bibl id="B110">
  <title><p>Determinants of undergraduate GPAs: SAT scores, high-school GPA and
  high-school rank</p></title>
  <aug>
    <au><snm>Cohn</snm><fnm>E</fnm></au>
    <au><snm>Cohn</snm><fnm>S</fnm></au>
    <au><snm>Balch</snm><fnm>DC</fnm></au>
    <au><snm>Bradley</snm><fnm>J</fnm></au>
  </aug>
  <source>Economics of Education Review</source>
  <publisher>Elsevier</publisher>
  <pubdate>2004</pubdate>
  <volume>23</volume>
  <issue>6</issue>
  <fpage>577</fpage>
  <lpage>-586</lpage>
</bibl>

<bibl id="B111">
  <title><p>The relation between socioeconomic status and academic
  achievement.</p></title>
  <aug>
    <au><snm>White</snm><fnm>KR</fnm></au>
  </aug>
  <source>Psychological bulletin</source>
  <publisher>American Psychological Association</publisher>
  <pubdate>1982</pubdate>
  <volume>91</volume>
  <issue>3</issue>
  <fpage>461</fpage>
</bibl>

<bibl id="B112">
  <title><p>Socioeconomic status and academic achievement: A meta-analytic
  review of research</p></title>
  <aug>
    <au><snm>Sirin</snm><fnm>SR</fnm></au>
  </aug>
  <source>Review of educational research</source>
  <publisher>Sage Publications Sage CA: Thousand Oaks, CA</publisher>
  <pubdate>2005</pubdate>
  <volume>75</volume>
  <issue>3</issue>
  <fpage>417</fpage>
  <lpage>-453</lpage>
</bibl>

<bibl id="B113">
  <title><p>Opportunity and Similarity in Dynamic Friendships</p></title>
  <aug>
    <au><snm>Bjerre Nielsen</snm><fnm>A</fnm></au>
    <au><snm>Dreyer Lassen</snm><fnm>D</fnm></au>
  </aug>
  <publisher>Unpublished manuscript</publisher>
  <pubdate>2017</pubdate>
</bibl>

<bibl id="B114">
  <title><p>Academic performance prediction in a gender-imbalanced
  environment</p></title>
  <aug>
    <au><snm>Sapiezynski</snm><fnm>P</fnm></au>
    <au><snm>Kassarnig</snm><fnm>V</fnm></au>
    <au><snm>Wilson</snm><fnm>C</fnm></au>
    <au><snm>Lehmann</snm><fnm>S</fnm></au>
    <au><snm>Mislove</snm><fnm>A</fnm></au>
  </aug>
  <source>FATREC Workshop on Responsible Recommendation Proceedings</source>
  <pubdate>2017</pubdate>
</bibl>

</refgrp>
} % end of \BMCxmlcomment

\newpage
\section*{Figures}

%%%%%%%%%%%%%%%%%%%%%%%%%%%%%%%%%%%
%%                               %%
%% Tables                        %%
%%                               %%
%%%%%%%%%%%%%%%%%%%%%%%%%%%%%%%%%%%

\newpage

\section*{Tables}

\begin{table}[!h]
\centering
\footnotesize 
	\begin{tabularx}{\columnwidth}[t]{ | Y |  Y | Y | Y | }
		%\hline
		\hline
		\textbf{Personality} & \textbf{Individual} & \textbf{Network} & \textbf{Combined}\\  \hline
		{\begin{tabularx}{.25\columnwidth}{@{\hskip-6pt}Y}
		BFI: Neuroticism\\
BFI: Openness\\
BFI:~Conscientiousness\\
BFI: Extraversion\\
BFI: Agreeableness \\
Satisfaction with Life \\ 
Locus of Control\\
PANAS: Positive\\
PANAS: Negative\\
Self-esteem\\
Loneliness\\
Stress\\
Depression\\
Narcissism: Rivalry\\
Narcissism:~Admiration\\
Narcissism: Overall\\
		\end{tabularx}} 
&
{\begin{tabularx}{.25\columnwidth}{@{\hskip-3pt}Y}
	Facebook activity\\
Class attendance\\
Gender\\
Study year\\
+ all personality features\\
		\end{tabularx}}

&
{\begin{tabularx}{.25\columnwidth}{@{\hskip-6pt}Y}
	\textbf{Calls}\\
Degree Centrality\\
Mean GPA of peers\\
Fraction of low/high performing peers\\
\\
	\textbf{Texts}\\
Degree Centrality\\
Mean GPA of peers\\
Fraction of low/high performing peers\\
\\
	\textbf{Proximity}\\
Degree Centrality\\
Mean GPA of peers\\
Fraction of low/high performing peers\\
\\
	\textbf{FB friends}\\
Degree Centrality\\
Mean GPA of peers\\
Fraction of low/high performing peers\\
\\
	\textbf{FB interactions}\\
Degree Centrality\\
Mean GPA of peers\\
Fraction of low/high performing peers\\
		\end{tabularx}}
&

All individual features and all network features together
\\

 \hline
	\end{tabularx}
	\par\smallskip
	\caption{Feature sets for data-driven modeling}
	\label{tab:feature_sets}
\end{table}

\begin{table}[!h]
	\small
	\centering
	\begin{tabularx}{.5\textwidth}{ | X  r | }
		\hline
		\textbf{Channel}  &  \textbf{$r_S$} \\
		\hline
		Texts & .432 \\
		Calls & .415 \\
		Facebook interactions & .323 \\
		Facebook friendships & .300 \\
		Proximity & .299 \\
		\hline
	\end{tabularx}
	\par\smallskip
	\caption{Correlation between the cumulative GPA of the students and the mean cumulative GPA of their peers based on different communication channels.
		Corresponding $p$-values are below $0.001$.}
	\label{tab:corr_GPA_friends}
\end{table}

\end{document}